\documentclass[journal,10pt]{IEEEtran}

\usepackage{graphics} 
\usepackage{epsfig} 
\usepackage{amsmath} 
\usepackage{amssymb}  
 \usepackage{csquotes}
\usepackage{bm}
\usepackage{amsmath}
\usepackage{url}
\input threeparttable.sty
\usepackage{gensymb}
\usepackage[utf8]{inputenc}
\usepackage[T1]{fontenc}
\usepackage{textcomp}
\usepackage{gensymb}
\usepackage{ragged2e}
\usepackage{algorithm}
\usepackage{algpseudocode}
\usepackage[noadjust]{cite}
\usepackage{color}
\begin{document}

\title{Structured Group Sparsity: A Novel Indoor WLAN Localization, Outlier Detection, and Radio Map Interpolation Scheme}

\author{Ali~Khalajmehrabadi,~\IEEEmembership{Student~Member,~IEEE,}
        Nikolaos~Gatsis,~\IEEEmembership{Member,~IEEE,}
        and~David~Akopian,~\IEEEmembership{Senior~Member,~IEEE}

\thanks{Manuscript received on March 6, 2016; revised May 9, 2016 and August, 10, 2016. The authors are with the Department of Electrical and Computer Engineering, The University of Texas at San Antonio, One UTSA Circle, San Antonio, Texas 78249-0669, USA.
 E-mails: ali.khalajmehrabadi@utsa.edu, nikolaos.gatsis@utsa.edu, david.akopian@utsa.edu}}

\markboth{IEEE Transactions on Vehicular Technology  (ACCEPTED WITH MINOR REVISIONS)}
{}

\maketitle

\begin{abstract}
This paper introduces novel schemes for indoor localization, outlier detection, and radio map interpolation using Wireless Local Area Networks (WLANs). The localization method consists of a novel multicomponent optimization technique that minimizes the squared $\ell_{2}$-norm of the residuals between the radio map and the online Received Signal Strength (RSS) measurements, the $\ell_{1}$-norm of the user's location vector, and weighted $\ell_{2}$-norms of layered groups of Reference Points (RPs). RPs are grouped using a new criterion based on the similarity between the so-called Access Point (AP) coverage vectors. In addition, since AP readings are prone to containing inordinate readings, called outliers, an augmented optimization problem is proposed to detect the outliers and localize the user with cleaned online measurements. Moreover, a novel scheme to record fingerprints from a smaller number of RPs and estimate the radio map at RPs without recorded fingerprints is developed using sparse recovery techniques.  All localization schemes are tested on RSS fingerprints collected from a real environment. The overall scheme has comparable complexity with competing approaches, while it performs with high accuracy under a small number of APs and finer granularity of RPs.
\end{abstract}


\IEEEpeerreviewmaketitle

\section{Introduction}
\IEEEPARstart{R}{ecently}, great attention has been directed toward providing Location Based Services (LBS) \cite{1} as the next generation of health care monitoring \cite{2,3} and content delivery services such as network management and security \cite{4}, personalized information delivery \cite{5}, and context awareness \cite{6}. Although these services are quite widespread for outdoor environments, challenges still exist indoors. These challenges primarily occur because there is no established localization-supporting infrastructure for indoor environments like the well-known Global Positioning System (GPS) for outdoors \cite{7}. 
\par Although the user's location can be enabled with the aid of proximity RF sensors \cite{58,r71}, existing RF infrastructures such as Bluetooth \cite{10}, infrared transceivers \cite{9}, visible light \cite{r81}, or even acoustic signals \cite{r82}, the cost of the necessary infrastructure for each building has prevented the proliferation of these approaches. Hence, attention has been directed toward existing infrastructures like Wireless Local Area Networks (WLANs) which are densely deployed indoors \cite{r212,r215,r62}. 
\par Modern WLAN indoor localization is performed through fingerprinting \cite{18,19,20,21,r39,r27,r33,54,55,r70,r71,r72,r73,r74,56,r195,r230,r234,r235,r237} which is a promising choice because WLAN signals are available in most indoor environments and RSS measurements are taken in most devices in the context of signal reception. In fingerprinting, the area is divided into several, usually equidistant, grid points also known as Reference Points (RPs). The localization approach has two phases: \textit{offline} and \textit{online}. In the offline phase, the Received Signal Strength (RSS) of the available Access Points (APs) is recorded at each RP. The set of \textit{fingerprints (location, RSS)} for the whole area is called the \textit{radio map} of the area. In the online phase, the user receives the RSS measurements from available APs and a rule, called \textit{localization scheme}, defines the relation between the online measurements and the radio map and estimates the user's location. In other words, the task of a fingerprint positioning system is to estimate the position of the user through a comparison mechanism between the online measurement and the radio map. 
\par WLAN localization faces several problems. In particular, the RSS fingerprints are prone to random fluctuations during the  fingerprinting phase, and inference of distance from the estimated attenuation is not accurate. The complex indoor environment renders a multipath RSS profile due to the presence of non-line-of-sight (NLOS) propagation with obstacles such as furnitures, doors, walls, etc. \cite{54,55}. Besides, RSS profiles show higher variance in short distances to APs \cite{56}.  
\par Motivated by the above problems, most of the modern approaches include the following four steps: 1) radio map pre-processing, where the radio map is processed in the offline phase to extract specific features from fingerprints, while clustering of RPs may also be performed \cite{r27,19}; 2)  coarse localization, in which a sub-region consisting of a subset of RPs that include the user's position is identified \cite{r27}; 3) AP selection, whereby a subset of available APs is chosen which can better differentiate between the RPs and introduces less biased readings \cite{r72,r68}; 4) fine localization where the user's location is estimated through a rule between the user's online measurements and radio map. 
\section{State-of-the-art WLAN Fingerprinting Localization: Related Work and Challenges}
\par In this section, main categories of fingerprinting localization approaches are reviewed followed by a discussion of existing challenges.
\subsection{Related Work}
\par The state-of-the-art localization approaches are divided into three broad categories mainly based on their fine localization methods: 1) probabilistic approaches which estimate the user's location based on statistical methods such as Kernel Density Estimation (KDE) \cite{19}, Conditional Random Fields (CRF) \cite{r195}, KL-divergence \cite{r230}, Dynamic Hybrid Projection
(DHP) \cite{r74}; 2) pattern recognition schemes such as Support-Vector Machines (SVM) \cite{r75,r76}; and 3) deterministic methods which rely on a distance metric between online measurements and radio map fingerprints. Deterministic approaches are divided into Nearest Neighbor (NN) \cite{8,r64,r77,r78}, geometry-based localization \cite{r234,r235,r237}, and Compressive Sensing (CS) methods \cite{r39,r27}.
\par The CS method formulates the localization as the solution to a set of linear equations. The radio map becomes the coefficient matrix, the user's online measurement is the observation vector, and user's location vector is the unknown vector to be estimated. The user can be only in one position at any single time and hence, the user's location vector is considered as a 1-sparse vector in which only the entry corresponding to the RP closest to the user location is 1 and the remaining entries are zero. The solution to a linear system of equations where the unknown vector is sparse can be obtained through an $\ell_{1}$-norm minimization rather than a simple pseudo-inverse solution. 
\par  WLAN localization methods can be enhanced by combining them with other positioning sources. For example, recent works reported such hybrid integrations with dead-reckoning, which is a method that uses the internal sensors of the hand-held devices, such as accelerometer, magnetometer, and gyroscope \cite{r83}. These methods exploit the spatial pattern of the fingerprints and fuse them with the the user's trajectory estimated from the sensors \cite{r177,r197,r199,r200}. While hybrid solutions are out of scope of this paper, the methods described  can be used in hybrid solutions as well.
\subsection{Challenges in Indoor WLAN Fingerprinting Localization: Outliers and Radio Map Construction}
\par Although CS-based localization is one of the most prominent fingerprinting positioning systems, it still suffers from several drawbacks. The CS-based method requires that the radio map matrix satisfies certain properties to render a unique solution \cite{31,32}. As a result, the radio map matrix needs an orthogonalization preprocessing, which may not work well numerically, because the radio map is a fat matrix. The orthogonalization step is also an extra step that increases computational complexity. Also, the CS approach assumes noiseless online RSS measurements and tries to find a position vector that matches the fingerprints exactly. However, online measurements cannot match exactly with the fingerprints of an RP. Furthermore, coarse localization should be applied before the fine localization step. This is an extra step and if it fails, it leads to a wrong subset of RPs and the entire localization fails \cite{r27,r33,r93}.
\par In addition, localization approaches may experience the unavailability of APs due to unforeseen reasons. For example, a set of APs may be lost due to natural disasters. In these cases, only a small number of APs usually remain functional. Online measurements deviate considerably from radio map fingerprints, and thus a localization system should be able to deliver the location of the user using only a subset of available APs.  WLANs are dynamic systems that experience rapid changes not only due to the AP software, but due to infrastructure changes such as removal of APs, malfunctions, jamming, intermittent shutting down of APs, or intentional adversary attacks that may weaken or strengthen the AP signals. In such cases, online readings of APs are not trustable. These inordinate readings are called \textit{outliers}, an issue that has surprisingly received little attention. 
\par Another challenge is related to the radio map construction and updating. On one hand, any changes in the environment such as AP relocation, structural changes, and equipment moving, change the characteristics of the environment. The radio map must be thus regularly updated. On the other hand, fine grained localization requires dense RP fingerprinting which takes a long time and incurs high labor costs. Therefore, some recent methods have been directed toward coarse fingerprinting and \emph{radio map interpolation} on a fine grid of RPs \cite{r27, r90,r91,r92}. Finally, while preventative measures including long-time fingerprinting, validation, and attack detection have been proposed before \cite{53}, the problem of radio map interpolation when the actual readings also contain outliers has not been addressed so far. 
\section{Contributions of this Paper}
The contributions of this paper are threefold and amount to novel schemes respectively for indoor WLAN localization, outlier detection, and radio map interpolation using (group) sparse recovery techniques.
\subsubsection{Group Sparsity-based Localization Method} 
This paper proposes a new WLAN fingerprinting localization approach that comprises three steps, in the online phase: 1) RP grouping; 2) AP selection; and 3) localization through group sparse regression. These steps and their merits are described next.
\par The first step entails computing the similarity between the online measurement and each RP fingerprint via an AP coverage vector. This similarity is the Hamming distance between the most available APs in the radio map. The most available APs are defined as those whose readings are above a specific threshold in most of the fingerprinting time. These similarities are used to group the RPs. These groups participate in localization through corresponding weights which are proportional to the inverse of the average group Hamming distance. 
\par In the second step, the number of APs engaged in localization is reduced through the conventional AP selection methods based on either the strongest set of APs \cite{18, r80}, which provide enlarged network coverage most of the time, or the Fisher Criterion \cite{18,20,46}. The latter is one of the most reliable AP selection methods and takes the fingerprinting history of the AP into account \cite{19}.
\par The last step is the user location estimation performed through Group-Sparsity (GS)-based regression. We introduce a model that better reflects the relation between the RSS fingerprints, online measurements, user's location, and noise, which models the components that cannot be explicitly defined.  The user location is estimated through the GS-based minimization which addresses the previously mentioned challenges and comprises a multi-objective optimization problem. The optimization chooses the RP fingerprints that minimize the difference between the radio map and online measurement, gives the opportunity to all RPs to engage in localization, and provides a sparse user's location. These objectives participate in the minimization problem through their corresponding tuning parameters.
\par Different from the CS-based approach, the new method has the following advantages: 1) There is no need for coarse localization as the two-step procedure of coarse localization followed by fine localization is not necessarily optimal. 2) Our solution utilizes all RPs in a single step localization, which does not suffer from the issue of potential error in coarse localization.  3) There is no need for preprocessing of the radio map (e.g., orthogonalization of the regression matrix). 4) The method is naturally tailored to noisy measurements, as the residuals are minimized instead of attempting to exactly match the online measurement to a fingerprint. 
\subsubsection{Outlier Detection}
A modified formulation of the GS-based localization system to account for outliers in the online phase is introduced. Basically, outliers model large errors in the online measurements or unavailable measurements from APs that were present in the fingerprinting phase. The main idea is that the outliers are explicitly modeled as an unknown vector, which may be efficiently recovered because it is sparse. Explicit modeling of outliers has been pursued in the statistics literature \cite{51}, but this is the first time that it is being incorporated in GS-based WLAN localization. Specifically, the new approach detects the APs with outliers, and performs the localization with clean online measurements. The location of the user and the outliers of APs are jointly estimated through a single optimization problem. 
\subsubsection{Radio Map Interpolation}
Motivated by the fact that fingerprinting is time consuming and incurs high labor cost, this paper develops an additional sparse recovery formulation that is able to interpolate between RPs. The radio map fingerprints can be recorded at a smaller number of RPs (coarser granularity) while finer granularity can be achieved through interpolation of fingerprints between RPs. We additionally tackle the problem of the radio map interpolation when readings contain outliers by augmenting the radio map interpolation scheme with an outlier detection component so that outliers have little effect on interpolation.

\par In summary, we formulate various aspects of WLAN localization leveraging group sparsity regression and optimization techniques. In the remainder of this paper, the offline radio map construction is described in Section \ref{Offline Phase:Radio map Construction}. The localization scheme and online RP layered clustering method are introduced in Section \ref{Online Phase}. Section \ref{Outlier Detection} describes the augmented localization model which jointly estimates the location and outliers. In Section \ref{Radio map Interpolation using Sparse Recovery}, the radio map interpolation scheme is developed. The experimental results of our work are detailed in Section \ref{Experiments and Discussion} followed by conclusions in Section \ref{Conclusion}.

\section{Offline Phase:Radio map Construction}
\label{Offline Phase:Radio map Construction}
For indoor WLAN fingerprinting localization, the area is divided into a set of RPs $ \mathcal{P}=\left\{\mathbf{p}_{j}=(x_{j},y_{j}), \  j=1,\ldots,N  \right\}$, where $  \mathbf{p}_{j}$ defines the RP Cartesian coordinates in $\mathbb{R}^{2}$. At each RP, RSS fingerprints are recorded at time instants $t_{m},\ m=1,\ldots,M$ with magnitudes $ r_{j}^{i}(t_{1}),\ldots, r_{j}^{i}(t_{M}) $, where $ i $ indicates the AP index from the set of APs $\mathcal{L}= \left\{AP^{1},\ldots,AP^{L}\right\} $. The RSS fingerprints from all APs at position $\mathbf{p}_{j}$ and at time $t_{m}$ are organized in a vector $ \mathbf{r}_{j}(t_{m})=[r_{j}^{1}(t_{m}),\ldots,r_{j}^{L}(t_{m})]^{T} $. The radio map at instant $t_{m}$ is the matrix consisting of the vectors $\mathbf{r}_{j}(t_{m}) , \ \ j=1, \ldots, N$, as columns. The time-averaged radio map,  $\mathbf{\Psi}_{L \times N}$, is defined as
\begin{equation}
\begin{split}
 \mathbf{\Psi}  =
 \begin{pmatrix}
   \psi_{1}^{1} & \cdots & \psi_{N}^{1} \\
   \vdots  &  \ddots & \vdots  \\
   \psi_{1}^{L} & \cdots & \psi_{N}^{L} 
  \end{pmatrix}\\
  \end{split}
\end{equation}
where $\psi _{j}^{i}=\frac{1}{M}\sum_{m=1}^{M} r_{j}^{i}(t_{m})$ and $ \boldsymbol \psi^{i}=[\psi_{1}^{i},\ldots ,\psi_{N}^{i}]$. The columns of radio map represent the average fingerprint readings at each RP from all APs. Since not all APs provide readings at each RP, the corresponding RSS value is set to a small value, $-95$ dBm, to imply its unavailability.
\section{Online Phase}
\label{Online Phase}
\begin{figure}[t!]
                       \includegraphics[scale=0.35]{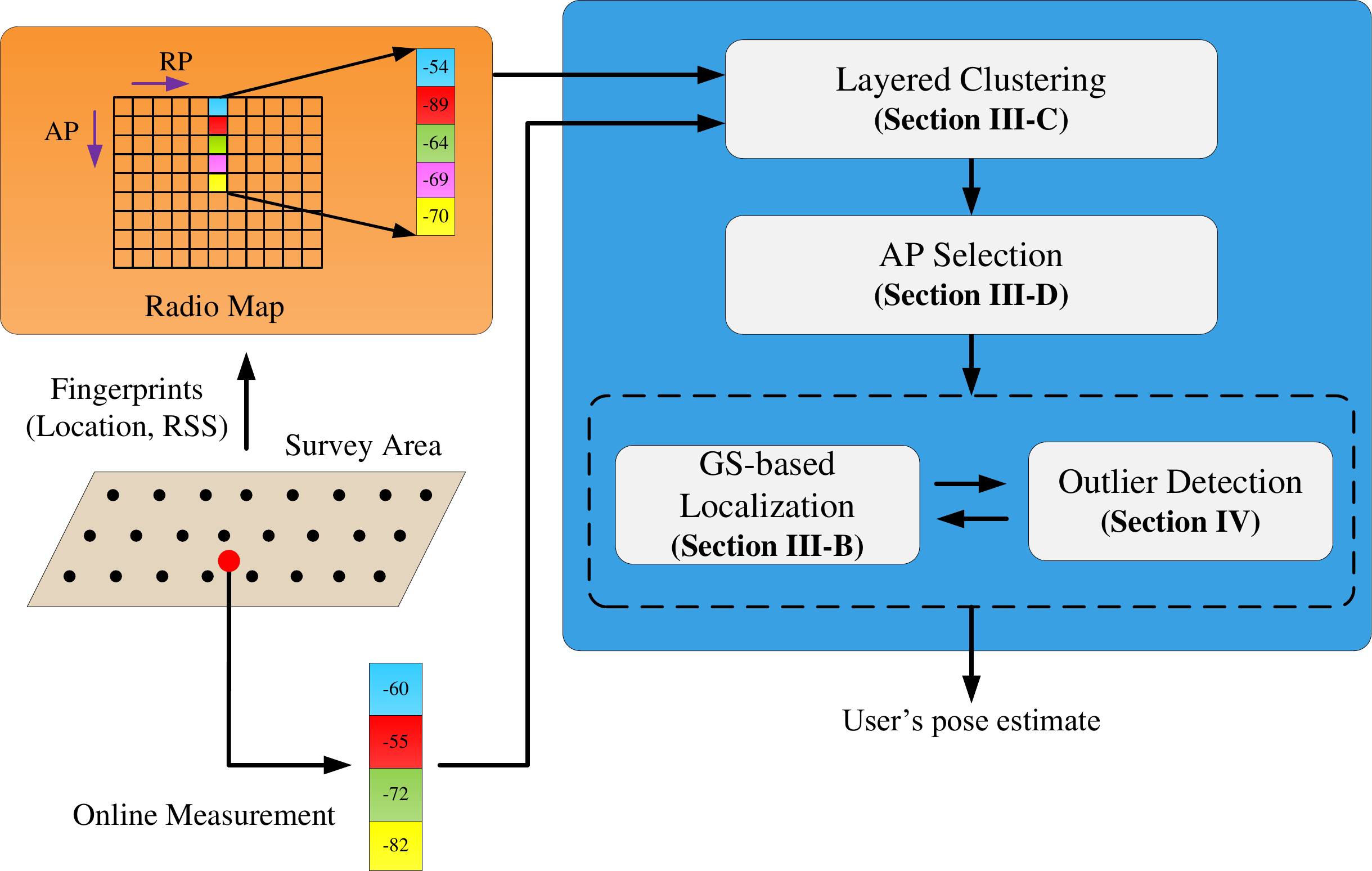}
                            \centering
                            \caption{The schematic of the proposed scheme.}
                            \label{figure6}
                         \end{figure}

Fig. \ref{figure6} shows the main tasks of our proposed sparse localization scheme. In the subsequent sections, we elaborate on its components. In this section, we first show how the WLAN fingerprinting localization problem can be formulated as a sparse recovery problem in Section \ref{Sparsity-based localization formulation}. To solve this problem, the efficient GS-based optimization is proposed in Section \ref{Group Sparsity-Based Localization}. Since this optimization problem divides the area into groups of RPs, an algorithm to cluster the RPs and compute their associated weights is developed in Section \ref{Online Layered Clustering}. The AP selection is the theme of Section \ref{AP Selection}. The localization with outlier detection is discussed in Section \ref{Outlier Detection}.
\subsection{Sparsity-based localization formulation}
\label{Sparsity-based localization formulation}
\par  In the online phase, the mobile user observes the measurements $ \boldsymbol y=(y^{1},\ldots,y^{L})^{T} $ at a single time.  A localization scheme uses the received online vector and the radio map $ \mathbf{\Psi}$ to estimate the mobile user's location, $  \hat{\mathbf{p}}= (\hat{x},\hat{y})$.
\par Since the user is in one location at every single time, the location of the user can be represented as a 1-sparse location indicator vector $\boldsymbol\theta = [ 0,\ldots,0,1,0,\ldots,0 ]^{T} $ where each entry corresponds to an RP and 1 shows the index of RP within which the user exists. Let also $ \mathbf{\Phi }$ be an $S \times L$ AP selection matrix where $S =\big |\mathcal{S} \big |, \ S < L$, $\mathcal{S} \subset \mathcal{L}$ and $\mathbf{y}$ be the vector of selected measurements that can be represented by
\begin{equation}
\label{eq0}
\mathbf{y}=\boldsymbol \Phi \boldsymbol y.
\end{equation}
Matrix $\boldsymbol \Phi$ has a single entry 1 at each row, corresponding to the selected AP. Particular methods to construct $ \mathbf{\Phi}$ will be outlined in Section \ref{AP Selection}. The sparsity-based model for WLAN localization problem can be represented as
\begin{equation}
\label{eq1}
\mathbf{y}=\mathbf{\Phi \Psi \boldsymbol\theta +\boldsymbol\epsilon}
\end{equation}
 where $ \boldsymbol \epsilon $ is a noise vector,
The model given by (3) and the fact that $\boldsymbol \theta$ is sparse will be the basis to recover $\boldsymbol \theta$ via sparse recovery techniques. 
\subsection{Group Sparsity-Based Localization} 
\label{Group Sparsity-Based Localization}
The main idea of the proposed localization scheme is to efficiently combine coarse and fine localization into a single step. The localization is thus performed by solving an optimization problem that combines the following objectives:
\begin{itemize}   
\item	Minimization of the squared $\ell_{2}$-norm of the residuals between the radio map and online measurements.
 \item	Recovery of a sparse vector of the user's location via minimization of the $\ell_{1}$-norm of the entire user's position vector.
           \item	Participation of all RPs in location estimation by accounting for the RP groups and their corresponding weights found in the first step (layered clustering). This is the group sparsity component and is expressed as a summation of the $\ell_{2}$-norms of groups of RPs. This component shrinks the non-zero elements of the position vector to a single group. 
          \end{itemize}
To this end, the GS-based localization is given by
\begin{equation}
  \label{eq3}
 \hat{\boldsymbol \theta} =\underset{\boldsymbol \theta}{\text{argmin} }\ \left[ \frac{1}{2}\Arrowvert \mathbf{y-\mathbf{H} \boldsymbol \theta}\Arrowvert_{2}^{2} + \lambda_{1} \Arrowvert\boldsymbol \theta\Arrowvert_{1}+\lambda_{2}\sum_{k=1}^{K}w_{k}\Arrowvert\boldsymbol \theta_{k}\Arrowvert_{2}\right]
  \end{equation}
where $ \mathbf{H}=\mathbf{\Phi \Psi }$, $\boldsymbol \theta_{k}$ is a part of location vector $\boldsymbol \theta$  that contains the indices of the RPs corresponding to group $k$, $w_{k}$ is the weight assigned to group $k$, $K$ is the total number of groups,  and $ \lambda_{1}, \lambda_{2} \ge 0$ are tuning parameters. The first argument minimizes the possible impact of online measurement noise considering that the RSS fingerprint noises have already been minimized through time-averaging of the fingerprints as described in Section \ref{Offline Phase:Radio map Construction}. The second component promotes sparsity in the position vector $\boldsymbol \theta$. The last term provides the sparsity over the groups (clusters) so that the recovered vector's nonzero elements are concentrated within a single group. This term basically plays the role of coarse localization. This minimization is known as Sparse Group Lasso (SGL) \cite{60}, and its custom solvers are available \cite{61}. It is applied to WLAN positioning in this paper for the first time.
\par The difference between the proposed GS-based and CS-based localization method is as follows:
\par 1) The CS approach does not follow model \eqref{eq1} and does not take the noise $\boldsymbol \epsilon$ into account. The GS is expected to have better performance in terms of localization accuracy because it considers the $\boldsymbol \epsilon$ component in the model. More specifically, the objective $\Arrowvert \mathbf{y-\mathbf{H} \boldsymbol \theta}\Arrowvert_{2}^{2}$ minimizes the difference between the radio map fingerprints and the online measurements.
\par 2) Another reason that group Lasso is expected to have better performance is that it is not restricted to choosing a subset of clusters, but all clusters participate in the fine localization. In other words, we have combined the coarse and fine localization stages into a single optimization.
\par 3) The coarse localization is an extra pre-processing step that increases the computational complexity of the whole localization approach. However, using all the clusters in GS method, there is no need for any coarse localization. 
\par 4) Another feature is the pre-processing steps that are required for the CS-based localization. In order to recover a unique sparse solution, the regression matrix needs to obey the Restricted Isometry Property (RIP) and mutual coherence conditions. For this reason, the radio map must be orthogonalized before solving the optimization problem which is not always workable as the matrix is not square. However, in our method, the solution takes advantage of any correlation between the radio map columns and does not need any pre-processing step.
\par Although formulation \eqref{eq3} is general and can use any subset of RP groups, a new clustering method is developed in the next subsection.
\subsection{Online Layered Clustering}
\label{Online Layered Clustering}
In this subsection, we propose a new method to define the groups of RPs and their corresponding weights. This method is summarized in Algorithm 1, and is described next in detail. 
\par First, define the AP coverage vector for the radio map as $\mathbf{I}_{j}=\left[I_{j}^{1},\ldots, I_{j}^{L} \right] $, where $I_{j}^{i}=1$ if AP $i$ provides continuous coverage at $\mathbf{p}_{j}$ and is $0$ otherwise. An AP provides  continuous coverage at $\mathbf{p}_{j}$ if its fingerprints are above a threshold $\gamma$ for 90 percent of the time samples \cite{19}. Similarly, for online vector $\boldsymbol y$, the coverage vector $\mathbf{I}_{\boldsymbol y}$ has its $i$-th entry set to 1 if the online measurement from AP $i$ is above $\gamma$, and zero otherwise. The Hamming distance between two binary vectors $\mathbf{I}_{\boldsymbol y}$ and $\mathbf{I}_{j}$ is used to indicate the number of APs with different coverage (lines 1-3 of Algorithm 1)
\begin{equation}
\label{eq2}
\begin{split}
d_{H}(\mathbf{I}_{\boldsymbol y}, \mathbf{I}_{j})= \sum _{i=1}^{L}|I_{y}^{i}-I_{j}^{i}|, \  j\in \{1,\ldots,N\}.
\end{split}
\end{equation}
\par For online layered clustering, the distance between the online measurement coverage vector and that of each RP is computed first using (\ref{eq2}). We define the minimum and maximum of the Hamming distance over the area as (lines 4, 5 of Algorithm 1)
\begin{equation}
\label{eq4}
\begin{split}
d_{H}^{min}&= \underset{j=1,\ldots,N}{\text{min}} \ d_{H}(\mathbf{I}_{\boldsymbol y}, \mathbf{I}_{j})\\
d_{H}^{max}&= \underset{j=1,\ldots,N}{\text{max}} \ d_{H}(\mathbf{I}_{\boldsymbol y},  \mathbf{I}_{j}).
\end{split}
\end{equation}
Then, the group Hamming range is defined, as follows
\begin{equation}
\label{eq5}
\begin{split}
r&=\frac{d_{H}^{max}-d_{H}^{min}}{K}
\end{split}
\end{equation}
where $K$ is the number of groups (clusters). RPs are clustered with respect to their Hamming distances to the online measurement. Specifically, the distance range $\left[ d_{H}^{min}, d_{H}^{max} \right] $ is partitioned in $K$ groups collected in set $\mathcal{D}$ (lines 6-8 of Algorithm 1)
\begin{equation}
\label{eq6}
\mathcal{D}=\left\lbrace  \left[ d_{k-1},d_{k}\right]  \big | d_{k}=d_{H}^{min}+(k-1)r, \ k=1,\ldots,K\right\rbrace .
\end{equation}
where $d_{0}=d_{H}^{min}$. Then, $\mathbf{p}_{j}$ is assigned to group $k$ if and only if (lines 10-13 of Algorithm 1)
\begin{equation}
\label{eq7}
d_{k-1}\le d_{H}(\mathbf{I}_{\boldsymbol y}, \mathbf{I}_{j}) \le d_{k}, \ \forall k=1,\ldots, K.
\end{equation}
Note that $\mathbf{p}_{j}$ cannot belong to more than one group. It could happen that $d_{H}(\mathbf{I}_{\boldsymbol y} , \mathbf{I}_{j})=d_{k} $, so, $\mathbf{p}_{j}$ may belong to groups $k$ and $k+1$. In this case, $\mathbf{p}_{j}$ is randomly assigned to one of these groups. The corresponding weight for each group is the inverse of the average of group Hamming distance (line 14 of Algorithm 1)
\begin{equation}
w_{k}=\frac{2}{d_{k-1}+d_{k} }, \ \forall k=1,\ldots, K.
\end{equation}

\begin{algorithm}[t!]
   \caption{: Layered Clustering}\label{a1}
   \begin{algorithmic}[1]
   \For{all $j \ \in \left\lbrace 1,\ldots , N \right\rbrace $}
   \State Compute  $ d_{H}(\mathbf{I}_{\boldsymbol y} , \mathbf{I}_{j}) $
   \EndFor
   \State Find  $d_{H}^{min}= \underset{j=1,\ldots,N}{\text{min}} \ d_{H}(\mathbf{I}_{\boldsymbol y}, \mathbf{I}_{j})$ and $d_{H}^{max}= \underset{j=1,\ldots,N}{\text{max}} \ d_{H}(\mathbf{I}_{\boldsymbol y},  \mathbf{I}_{j})$
   \State Compute $ r=\frac{d_{H}^{max}-d_{H}^{min}}{K}$
   \For {$k=1,\ldots, K$}
   \State Define $\mathcal{D}_{k}= [d_{k-1},d_{k}]$ where $ d_{k}=d_{H}^{min}+(k-1)r$ for $k=1,\ldots,K$ and $ d_{0}=d_{H}^{min}$
   \EndFor
   \State $\mathcal{C}_{k}=\varnothing, \ \forall k=1,\ldots, K$
   \For{all RPs  $\mathbf{p}_{j}$ }
   \State Find $k$ where $d_{k-1}\le d_{H}(\mathbf{I}_{\boldsymbol y}, \mathbf{I}_{j}) \le d_{k} \ \ \forall k=1,\ldots, K.$
   \State  $\mathcal{C}_{k}=\left\lbrace \mathcal{C}_{k},\mathbf{p}_{j}\right\rbrace  $
   \EndFor
   \State $w_{k}=\frac{2}{d_{k-1}+d_{k} }, \ \forall k=1,\ldots, K$
   \end{algorithmic}
   \end{algorithm}
Then, the groups with their corresponding weights participate in optimization \eqref{eq3}.
\subsection{AP Selection}
\label{AP Selection}
\par Since WLANs are designed to provide the maximum network coverage over the area, the number of available APs in an area is much greater than that required for positioning. Also, not all APs are suitable for positioning as the ones with high variance introduce large errors and lead to biased estimations. This motivates us to engage a set of APs in the localization that can efficiently represent the characteristics of the environment.
\par In our work, we have considered two AP selection methods:
\begin{itemize}   
\item	\textit{Strongest APs Selection}: In this method, we consider the set of APs whose online readings are above a predefined threshold. The intuition behind this idea is that the strongest APs can provide the best features of the area. For this, the entire of the online measurement vector are sorted in decreasing order and $ \left|  \mathcal{S} \right|$ strongest APs are selected where $\mathcal{S}\subset \mathcal{A} $.  This method works well if the characteristics of the environment have not been changed from the fingerprinting time. Our results in Section \ref{Radio map Interpolation using Sparse Recovery} show that this assumption is not valid if the fingerprinting and localization times are distant. 
\item	\textit{Fisher criterion}: This criterion uses the history of radio map fingerprints and does not integrate the online measurements in selecting APs. This criterion assigns a score to an AP which is proportional to the differentiability of APs across RPs and inversely proportional to the variance of readings for that AP through the fingerprinting time:
 \begin{equation}
  \begin{split}
 \zeta^{i}&=\frac{\sum_{j=1}^{N}( \psi_{j}^{i}-\bar{ \psi}^{i})^{2}}{\frac{1}{M-1}\sum_{m=1}^{M}\sum_{j=1}^{N}(r_{j}^{i}(t_{m})- \psi_{j}^{i})^2}, \
  i=1,\ldots,L
  \end{split}
  \end{equation}
  where 
  \begin{equation}
  \bar{ \psi}^{i}= \frac{1}{N}\sum_{j=1}^{N} { \psi}_{j}^{i} 
  \end{equation} 
  For AP selection, the Fisher score is sorted decreasingly and a subset of $ \left|  \mathcal{S} \right|, \ \mathcal{S} \subset \mathcal{A} $, APs with the greatest Fisher score are selected.
          \end{itemize}
\par In online localization, the AP selection procedure is modeled by a matrix $\mathbf{\Phi}$ whose $i$-th row, $\mathbf{\Phi}^{i}$, defines the AP that is selected. Each row of $\mathbf{\Phi}$ is a binary 1-sparse $1 \times N $ vector whose only $i$-th index is 1, indicating the selected $i$-th AP as 
\begin{equation}
  \mathbf{ \Phi}^{i}=[0,\ldots,\underbrace{1}_\text{Index of selected AP},\ldots,0] \ \ \ \ \ \forall i=1,\ldots, S.
  \end{equation}
\par This AP selection matrix is used in the online measurement equations \eqref{eq0} and \eqref{eq1}.
\par Finally, the location of the mobile user is computed as the centroid of the convex hull generated by the RPs as
\begin{equation}
   \hat{\mathbf{p}}=(\hat{x},\hat{y})=\frac{1}{\sum_{ j=1}^{N}\hat{\theta}_{j}} \sum_{j=1}^{N}\hat{\theta}_{j}\cdot \mathbf{p}_{j}.
   \end{equation}
\section{Outlier Detection}
\label{Outlier Detection}
The previous section elaborated on the main features of the proposed localization scheme. In this section, the issue of localization with the presence of outliers is addressed. Here, we only consider the outlier detection scheme in the online phase. The radio map outlier detection for interpolated fingerprints is discussed in Section \ref{Interpolation with Radio Map Outliers}.  
\par Outliers cannot be detected by AP selection methods as they are mainly focused on selecting APs for the best differentiability between RPs. If AP readings contain outliers, penalizing $\ell_{1}$-norm of location vector alone does not guarantee outlier rejection and positioning suffers dramatically from large localization errors.
 
\par According to the above discussion, outliers can be integrated in our previous sparsity-based localization model \eqref{eq1} as
\begin{equation}
\label{eq24}
\mathbf{y}=\mathbf{\Phi \Psi \boldsymbol\theta +\boldsymbol \kappa+\boldsymbol\epsilon}
\end{equation}
where $\boldsymbol \kappa$ is a vector indicating the outliers and is sparse because the number of APs that may contain outliers is small. Hence, $\boldsymbol \kappa$ can be jointly estimated with the user location vector using an $\ell_{1}$-norm regularization. 
\par The modified Group-Sparsity (MGS)-based minimization of \eqref{eq3} can be represented as
\begin{equation}
\label{eq8}
\begin{split}
 (\hat{\boldsymbol \theta},\hat{\boldsymbol \kappa}) =\underset{\boldsymbol \theta, \boldsymbol \kappa}{\text{argmin} }\ &\left[ \frac{1}{2}\Arrowvert \mathbf{y-\mathbf{H}\boldsymbol \theta-\boldsymbol \kappa }\Arrowvert_{2}^{2} + \lambda_{1} \Arrowvert\boldsymbol \theta\Arrowvert_{1}+P_{\alpha}\right]\\
 & P_{\alpha}=\lambda_{2}\sum_{k=1}^{K}w_{k}\Arrowvert\boldsymbol \theta_{k}\Arrowvert_{2}+\lambda_{3}\Arrowvert \boldsymbol \kappa \Arrowvert_{1}.
\end{split}  
  \end{equation}
The outlier vector $\boldsymbol \kappa$ enables the optimization \eqref{eq8} to discard the outliers and find the user's location with the cleaned measurements. In essence, $\boldsymbol \kappa$ absorbs significant deviations from fingerprints. Optimization \eqref{eq8} amounts to a second order cone program, which can be efficiently solved \cite{cvx,gb08}.
  \begin{figure}[t!]
         \includegraphics[scale=0.5]{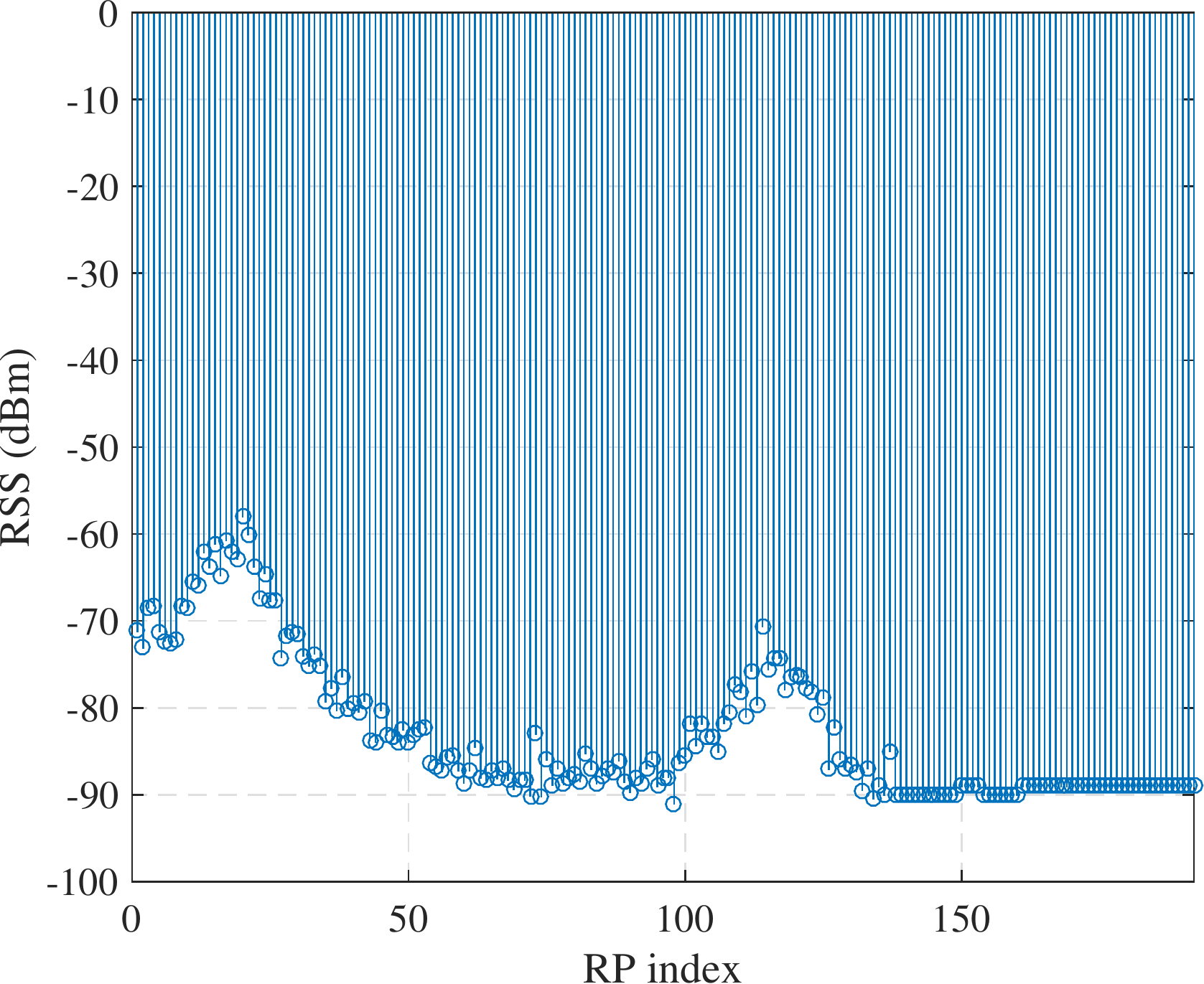}
         \centering
         \caption{The RSS profile for a single AP through the whole area over 192 RPs. The smoothness of the profile shows a great potential for interpolation.}
          \label{figure7}    
  \end{figure}
\section{Radio map Interpolation using Sparse Recovery}
\label{Radio map Interpolation using Sparse Recovery}
In this section, we first formulate the interpolation of radio map using its sparse frequency equivalent and introduce our interpolation scheme. Then, we discuss the interpolation of the radio map if the measured fingerprints also contain outliers. 
 \begin{figure}[t!]
        \includegraphics[scale=0.505]{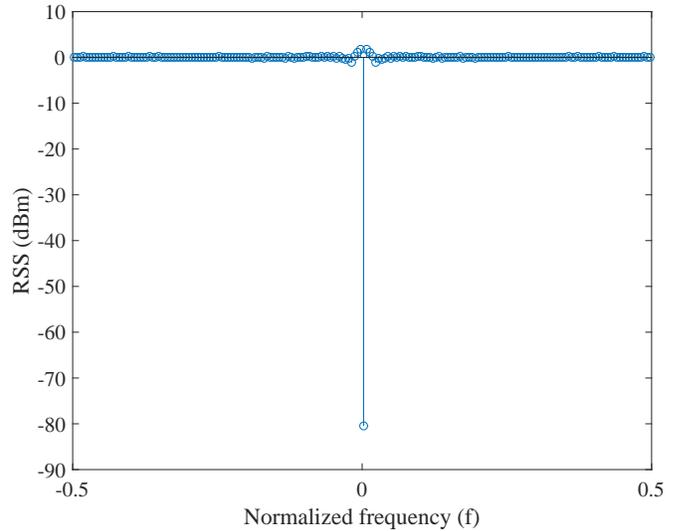}
        \centering
        \caption{The frequency representation of the RSS fingerprints for an AP. The great sparse profile shows the potential for RSS interpolation using the sparse recover algorithm.}
         \label{figure10}    
 \end{figure}
\subsection{Interpolation Procedure Formulation}
The RSS fingerprints can be interpolated at some RPs if we measure the radio map at other RPs. In other words, the radio map fingerprints can be collected at coarser granularity and the RSS fingerprints can be estimated at a denser granularity through interpolating the radio map between RPs. Fig. \ref{figure7} shows the RSS profile for a single AP over 192 RPs. This shows the smoothness of the profile and a suitable potential for interpolating the magnitudes of some RPs from their adjacent RPs. Fig. \ref{figure10} also represents the frequency representation of the RSS profile. This representation shows the very sparse feature of fingerprints which can be utilized for sparse reconstruction (interpolation). 
\par The sparse recovery can also be used in the offline phase to reconstruct the radio map from a lower number of RSS fingerprints. Let $\mathbf{F}$ be the $N\times N$  Fourier transform matrix that linearly transforms the vector of radio map fingerprints to its equivalent representation in the frequency domain as
\begin{equation}
\label{eq17}
\boldsymbol \psi^{i} _{f}=\mathbf{F}\boldsymbol \psi^{i}, \ i=1,\ldots,L.
\end{equation}
The representation of $\boldsymbol \psi^{i} _{f}$ is sparse which means most of the frequency components are zero---see e.g., \cite{r27}---and our empirical evidence also confirms it. This observation helps us reconstruct the radio map in the subsequent discussions. Then, consider a matrix that defines the relation between all RPs and the ones that we have taken the fingerprints over. To this end, we define a matrix $\mathbf{A}_{V\times N}$ whose rows are 1-sparse vectors $\mathbf{a}^{i}=[0,\ldots,1,\ldots,0]$ denoting the index of the RP that is measured during radio map fingerprinting, and $V$ is the total number of these RPs. In essence, $\mathbf{A}$ selects the RPs in which we record actual fingerprints.
\par The model for the offline radio map interpolation for AP $i$  can be represented as
\begin{equation}
\label{eq18}
\begin{split}
\mathbf{b}^{i}=\mathbf{A}\boldsymbol \psi^{i}=\mathbf{A}\mathbf{F}^{-1}\boldsymbol \psi^{i} _{f}, \  i=1,\ldots,L.
\end{split}
\end{equation}
Equation \eqref{eq18} is an under-determined system of equations because $V<N$. However, since $\boldsymbol \psi^{i} _{f}$ is sparse, a unique solution exists for it. To find the unique solution for \eqref{eq18}, we propose a special case of the group sparse recovery formulation as
\begin{equation}
  \label{eq19}
 \hat{\boldsymbol \psi}^{i} _{f} =\underset{\boldsymbol \psi^{i} _{f}}{\text{argmin} }\ \left[ \frac{1}{2}\Arrowvert \mathbf{b}^{i}-\mathbf{A}\mathbf{F}^{-1}\boldsymbol \psi^{i} _{f}\Arrowvert_{2}^{2} + \lambda_{1} \Arrowvert \boldsymbol \psi^{i} _{f}\Arrowvert_{1}\right]
  \end{equation}
 which has the form of the group sparse recovery \eqref{eq3} with $\lambda_{2}=0$. This reformulated minimization problem is known as Least Absolute Shrinkage and Selection Operator (LASSO). The above formulation minimizes the error between the measured RSS fingerprints and the interpolated fingerprints, while the second term provokes sparsity of  the RSS fingerprints in the Fourier domain. 
 \par The previous optimization is solved for all APs. The reconstructed radio map rows are computed as
 \begin{equation}
 \hat{\boldsymbol \psi}^{i}=\mathbf{F}^{-1}\hat{\boldsymbol \psi}^{i}_{f}
 \end{equation}
 where $ \hat{\boldsymbol \psi}^{i}$ is the reconstructed radio map for AP $i$. Using \eqref{eq19}, RSS fingerprints can be measured on a smaller number of RPs, and the radio map is interpolated in between RPs for a finer granularity.
 \par In the interpolation, some RPs are skipped and choosing a selection order for faithful radio map reconstruction is a problem discussed in Appendix. For the rest of this paper, the interpolation will be used on random selection of RPs.
 \subsection{Interpolation with Radio Map Outliers}
 \label{Interpolation with Radio Map Outliers}
 \par In Section \ref{Outlier Detection}, we discussed the outlier detection for the user's online measurement vector. The outliers may also happen in the offline phase for radio map fingerprints. If the radio map is interpolated in the offline phase and the APs whose RSS fingerprints are recorded contain outliers at specific RPs, the interpolation procedure should also contain an outlier detection scheme. The extension of \eqref{eq18} to include outliers is as follows:
 \begin{equation}
 \label{eq21}
 \begin{split}
 \mathbf{b}^{i}=\mathbf{A} \boldsymbol \psi^{i}+\boldsymbol \kappa=\mathbf{A}\mathbf{F}^{-1}\boldsymbol \psi^{i}_{f} + \boldsymbol \kappa, \ i=1,\ldots,L.
 \end{split}
 \end{equation}
 Based on the previous model, the modified radio map reconstruction scheme which contains the outlier detection component is
 \begin{equation}
   \label{eq20}
   \begin{split}
   (\hat{\boldsymbol \psi}^{i} _{f},\hat{\boldsymbol \kappa}) =& \ \underset{ \boldsymbol \psi^{i} _{f},\boldsymbol \kappa }{\text{argmin} }\ \left[ \frac{1}{2}\Arrowvert \mathbf{b}^{i}-\mathbf{A}\mathbf{F}^{-1}\boldsymbol \psi^{i} _{f}-\boldsymbol \kappa\Arrowvert_{2}^{2} + P_{\alpha}\right]\\
  &P_{\alpha}=\lambda_{1} \Arrowvert \boldsymbol \psi^{i} _{f}\Arrowvert_{1}+\lambda_{2} \Arrowvert \boldsymbol \kappa \Arrowvert_{1}.
   \end{split}
   \end{equation}
The above minimization reduces the effect of outlier for AP $i$ by detecting the outlier, while interpolating between RPs with clean measurements.
\begin{figure*}[t!]
                    \includegraphics[scale=0.945,angle =90]{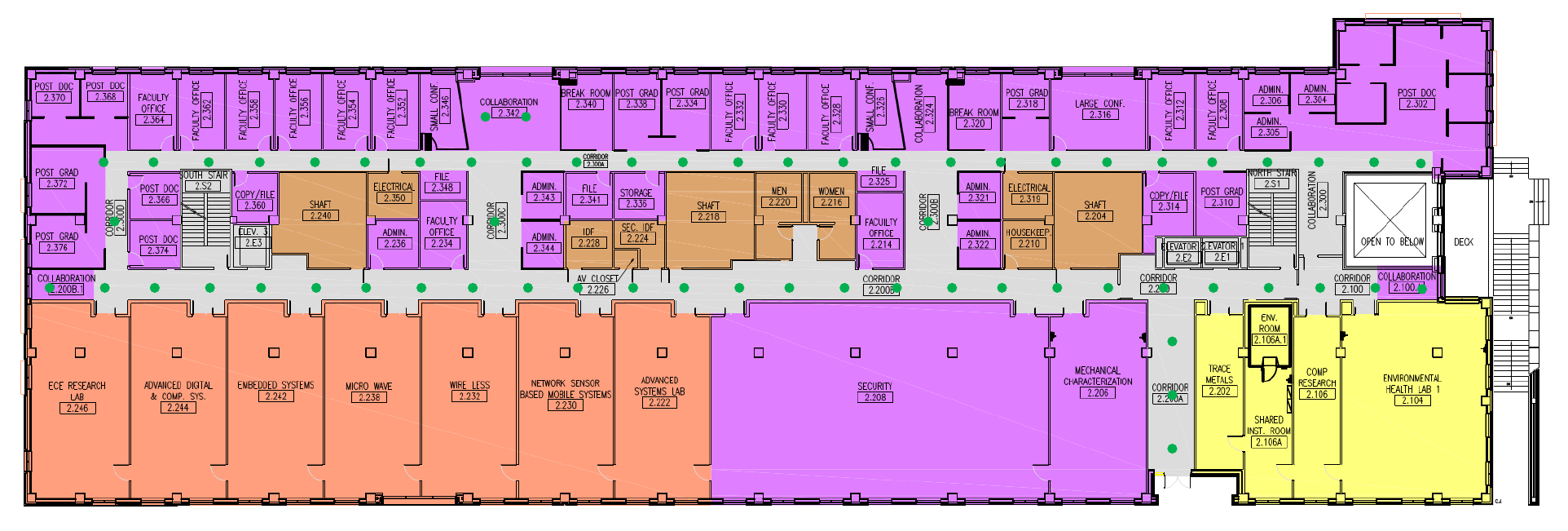}
                         \centering
                         \caption{The map of experimental environment. The green dots indicate the RPs.}
                         \label{fig2}
                      \end{figure*}
\section{Experiments and Discussion}
\label{Experiments and Discussion}
In this section, we present our experimental results that confirm the advantages of the proposed approach. The experimental setup, evaluation criteria, and the results are elaborated in sequel.

\subsection {Fingerprinting Device}
A collection of fingerprints have been recorded using a Samsung Tablet (Galaxy Tab A) and a self-developed Android application functioning on Android Lollipop 5.0.2 and utilizing the inherent android.net.wifi package features. This package gives the freedom to set the reading interval of the device Network Interface Card (NIC) to desired values. To the best of our knowledge, the device has a delay in updating the NIC card and thus, the sampling intervals cannot be set to small values. We set the application to read the NIC once per second to ensure obtaining the updated RSS values. The device also records the Media Access Control (MAC) address of each AP along with its RSS magnitude.

\subsection {Fingerprinting Setup}
\par The fingerprints have been collected from the second floor of the Applied Engineering Technology (AET) building with an area of $576 \ \mathrm{ft} \times 35 \ \mathrm{ft}$, and Biotechnology Sciences and Engineering (BSE) building with an area of $114 \ \mathrm{ft} \times 347 \ \mathrm{ft}$, both at the University of Texas at San Antonio  which are comparable to those reported in \cite{19,r27}. Fig. \ref{fig2} shows the map of the AET second floor in which the green dots depict the fingerprinting locations. The building features a representative indoor office environment with complex wireless propagation patterns due to research labs, offices, a library, and study areas, and has a high commuting volume. The BSE floor is structurally different from that of AET as it contains labs, offices, conference rooms, wide open hall, and a cafeteria. 
\par The WiFi signals lack the ability to exactly differentiate between points. So, a dense granularity leads to unnecessary and inefficient redundancy while a wide granularity leads to a low localization precision. Although existing methods set the RP spacing to 5-9 ft, we found 3 ft to be a sound grid spacing.  
\par During fingerprinting, a total of $L=268$ different MAC addresses were visible at AET while the device could read $L=1238$ different MAC addresses at BSE building. The MAC addresses were used as the unique identifiers of available APs. The fingerprints have been collected over three weeks during office hours to be representative of the APs RSS variations. 
\par The localization results in this section have been gathered through evaluating our algorithms in $N_{t}=100$ different test points with random positions. 
\subsection {Evaluation Metrics}
The figure-of-merit for evaluating the method accuracy is the average of the euclidean distance between the estimated and the true locations known as Mean Absolute Error (MAE) \cite{r39,r27,19,r72}
   \begin{equation}
   \text{MAE}=\frac{1}{N_{t}}\sum_{n=1}^{N_{t}}\sqrt{ \left( \hat{\mathbf{p}}(n)-\mathbf{p}(n)\right) ^{T}\left( \hat{\mathbf{p}}(n)-\mathbf{p}(n)\right) } 
   \end{equation}  
where $N_{t}$ is the number of test points. 
\par Other indicators of the performance of a localization system include the minimum and maximum errors, and frequency of errors. A suitable metric that includes these features is the empirical Cumulative Distribution Function (CDF) of errors.
\subsection {Experimental Results of GS-based  Localization Approach}
\par The localization performance is dependent on the AP selection scheme. Fig. 3 depicts the average localization errors under the two different AP selection methods, namely the strongest AP selection and Fisher criterion when the area has been divided into $K=15$ groups. The horizontal axis shows the number of APs used for localization. The AP selection based on Fisher criterion has definitely better performance. For the rest of results, the Fisher criterion has been chosen as the AP selection mechanism.
\par Fig. 4 shows the MAE of CS-based and GS-based localization for an increasing number of APs. In implementing the CS-based localization, the whole radio map has been used without utilizing any coarse localization. The localization error for the CS approach is large if the number of APs is small. The intuition behind this performance is that a large number of APs $(S>logN)$ is required  to render unique recovery of the position vector \cite{31,32}. For a better performance with a low number of APs, the CS-based localization needs to resort to coarse localization to decrease the number of RPs, $N$. However, the performance of GS-based localization remains quite constant regardless of the number of utilized APs and is slightly enhanced when more APs are engaged in localization. The least localization error is achieved when 12 APs are used.

\begin{figure}[t!]
  \label{fig11}
         \includegraphics[scale=0.49]{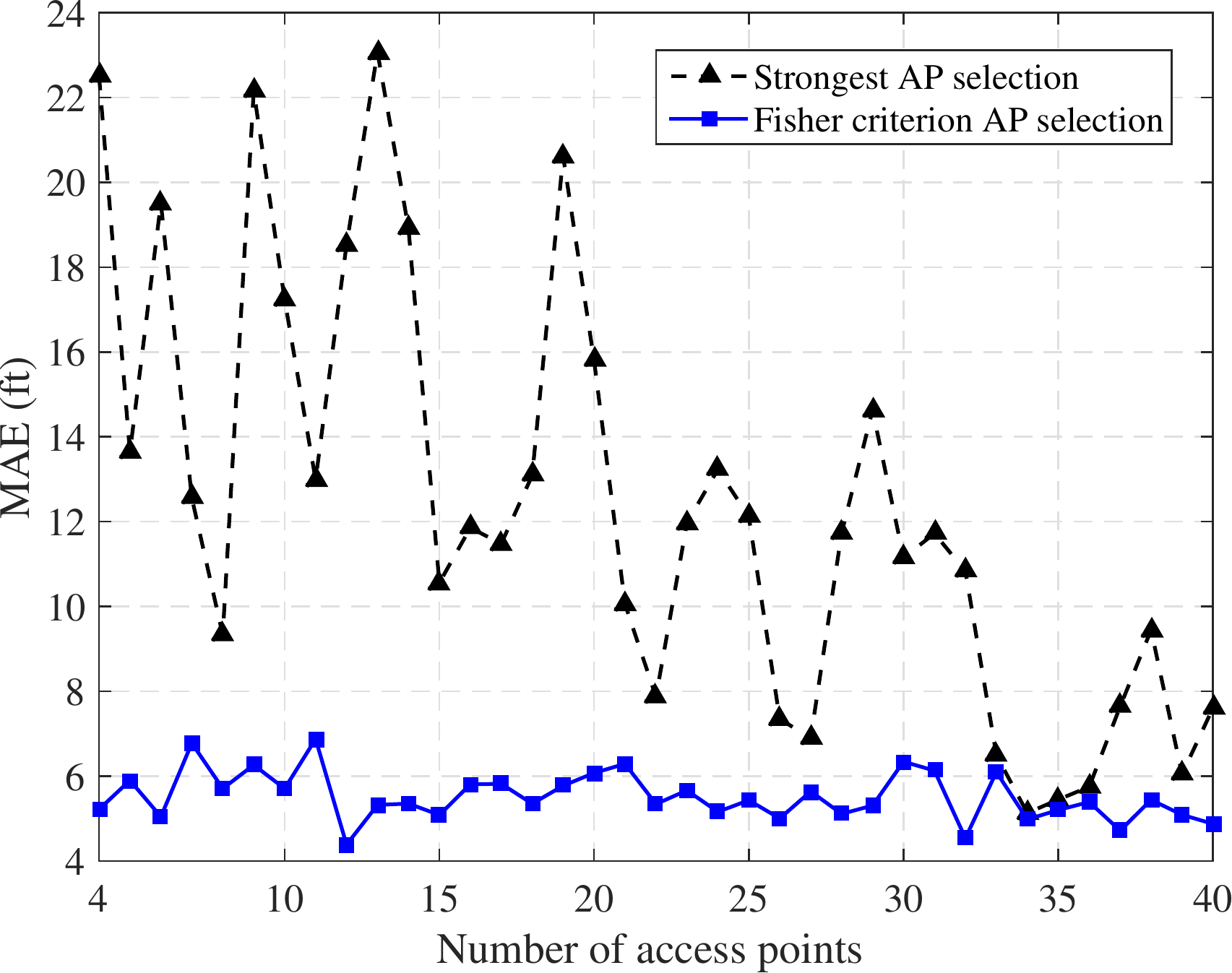}
        
            \centering      \caption{The MAE for GS-based localization using the two different AP selection schemes: Strongest AP and Fisher Criteron.}
                             
\end{figure}                                                
 \begin{figure}[t!]
         \includegraphics[scale=0.49]{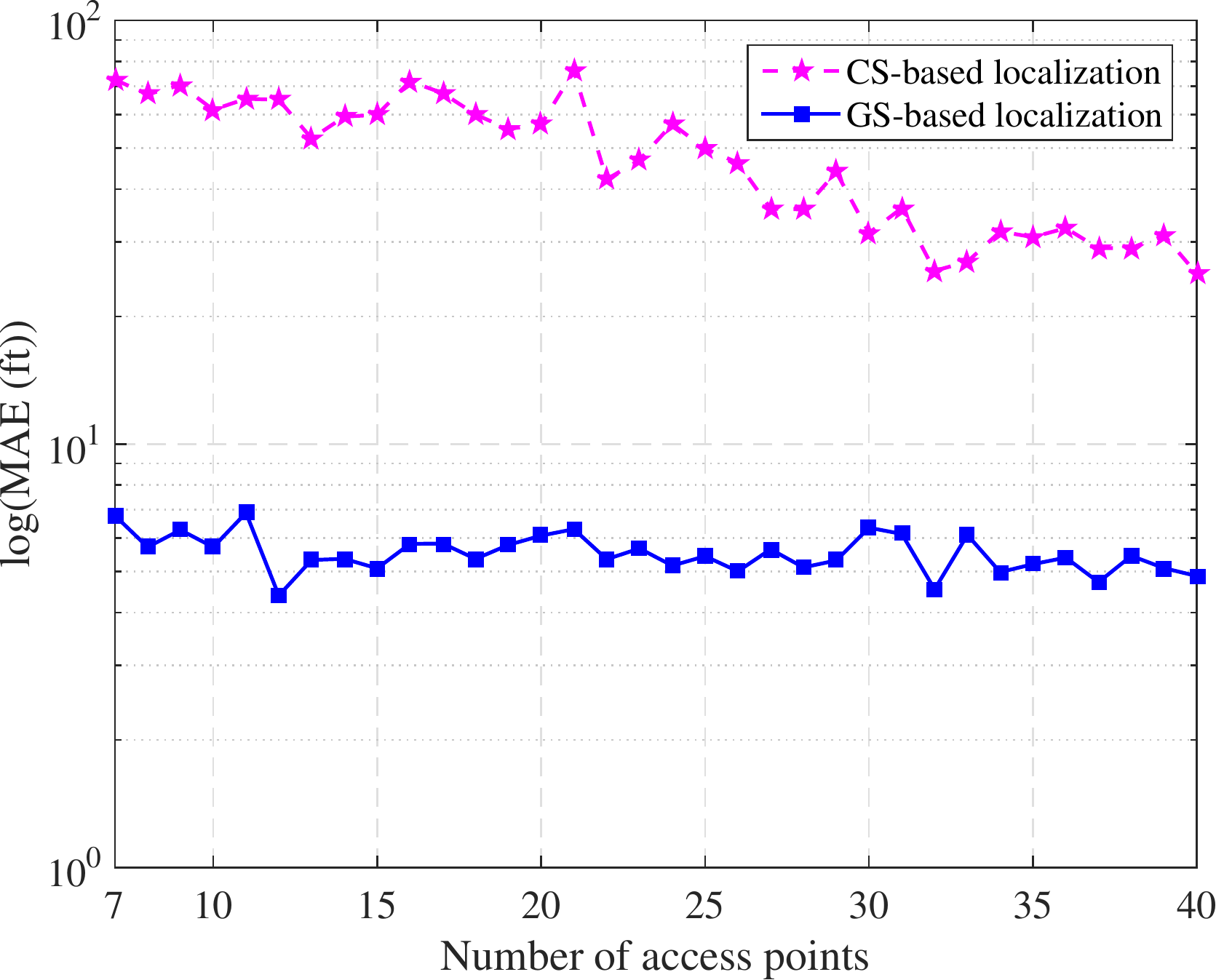}
          \label{fig4}
            \centering      \caption{The MAE for GS and CS-based localization with respect to different number of access points. The CS error decreases when more AP measurements are used for localization. The GS-based localization accuracy slightly increases if more APs are used.}
                             
                           \end{figure}         
              \begin{figure}[t!]
                       \includegraphics[scale=0.48]{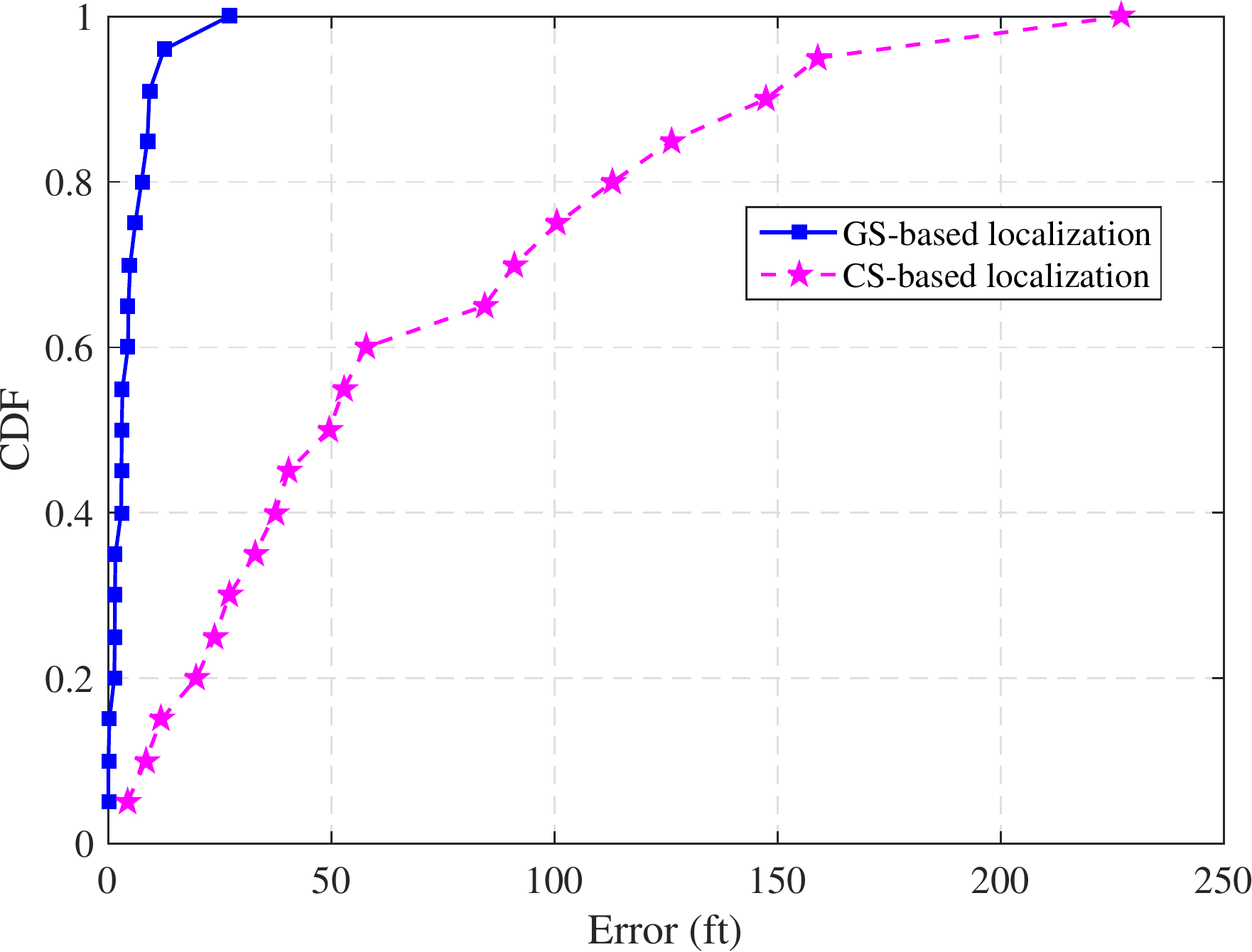}
                            \centering
                            \caption{The cumulative position error for GS and CS-based Localization. The results show that CS fundamentally depends on the coarse localization scheme and the error would be large otherwise.}
                            \label{figure4}
                         \end{figure}
\par The cumulative distribution of errors of the proposed approach is compared against that of the CS-based localization and depicted in Fig. \ref{figure4}. We observe large errors in CS-based localization which indicates that the solution is misled. The proposed GS-based localization shows very desired CDF characteristics as most of the errors are concentrated below 10 ft with the maximum at 22ft.
\par The localization error statistics of the GS-based approach is also compared against those of some of the recent approaches in Table \ref{table2}. The Centaur is an interesting combination of WiFi and acoustic ranging (AR) localization \cite{r86}. The table shows the the GS has smaller errors than Centaur.
\par The statistics of the localization error at the AET and BSE environment is illustrated in Table \ref{buildingcompare} for $L=4$ and $L=10$ APs. The error statistics are comparable which shows the consistency of the method over different environments.
\begin{table}[t!]
                      \caption{Position Error Statistics Comparison}
                       \label{table2}
                       \label{compare}
                      \begin{center}
                      \begin{tabular}{ |c|c|c|c|c| } 
                      \hline
                      \textbf{Methods} & \textbf{25\% (ft})& \textbf{50\% (ft)}& \textbf{75\%(ft)} & \textbf{100\% (ft)}\\
                      \hline
                      KNN-based \cite{8} & 6.2 & 9.6 & 15.3 & 72.1  \\ 
                        \hline
                      Kernel-based \cite{19} & 3.2 & 6.5 & 9.8 & 39.3 \\ 
                        \hline
                      Centaur \cite{r85} & 3.2 & 6.2 & 9.5 & 32.8 \\ 
                        \hline
                      Tilejunction \cite{r234} & 12.7 & 19.6 & 26.2& 68.8\\
                        \hline
                      Contour-based \cite{r237} & 7.2 & 10.4 & 18 & 52.4\\
                        \hline
                      GS-based& 1.2 & 4.1 & 8.6 & 22\\
 
                      \hline
                      \end{tabular}
                      \end{center}
                       \end{table}
 \begin{table*}[t!]
                       \caption{Position Error Statistics Comparison for AET and BSE Environments}
                        \label{table3}
                        \label{buildingcompare}
                       \begin{center}
                       \begin{tabular}{ |c|c|c|c|c| } 
                       \hline
                       \textbf{Percentages} & \begin{tabular}[x]{@{}c@{}} \textbf{AET}\\\textbf{L=4, K=15}\end{tabular}  & \begin{tabular}[x]{@{}c@{}} \textbf{AET}\\\textbf{L=10, K=15}\end{tabular}  & \begin{tabular}[x]{@{}c@{}} \textbf{BSE}\\\textbf{L=4, K=25}\end{tabular} & \begin{tabular}[x]{@{}c@{}} \textbf{BSE}\\\textbf{L=10, K=25}\end{tabular}\\
                       \hline
                       \textbf{25\% (ft})  & 1.49 & 1.51 & 0 & 1.2  \\ 
                         \hline
                       \textbf{50\% (ft)} & 3.17 & 3.28 & 3 & 3 \\ 
                         \hline
                       \textbf{75\%(ft)} & 7.41 & 7.1 & 7.5 & 6.29 \\ 
                         \hline
                       \textbf{100\% (ft)}& 29.26 & 17.2 & 39.3 & 33 \\
                         \hline
                         \textbf{Avg (ft)}& 5.6 & 5.74 & 7.69 & 4.97 \\
                                                  \hline
                       \end{tabular}
                       \end{center}
                        \end{table*}
         \begin{figure}[t!]
                       \includegraphics[scale=0.53]{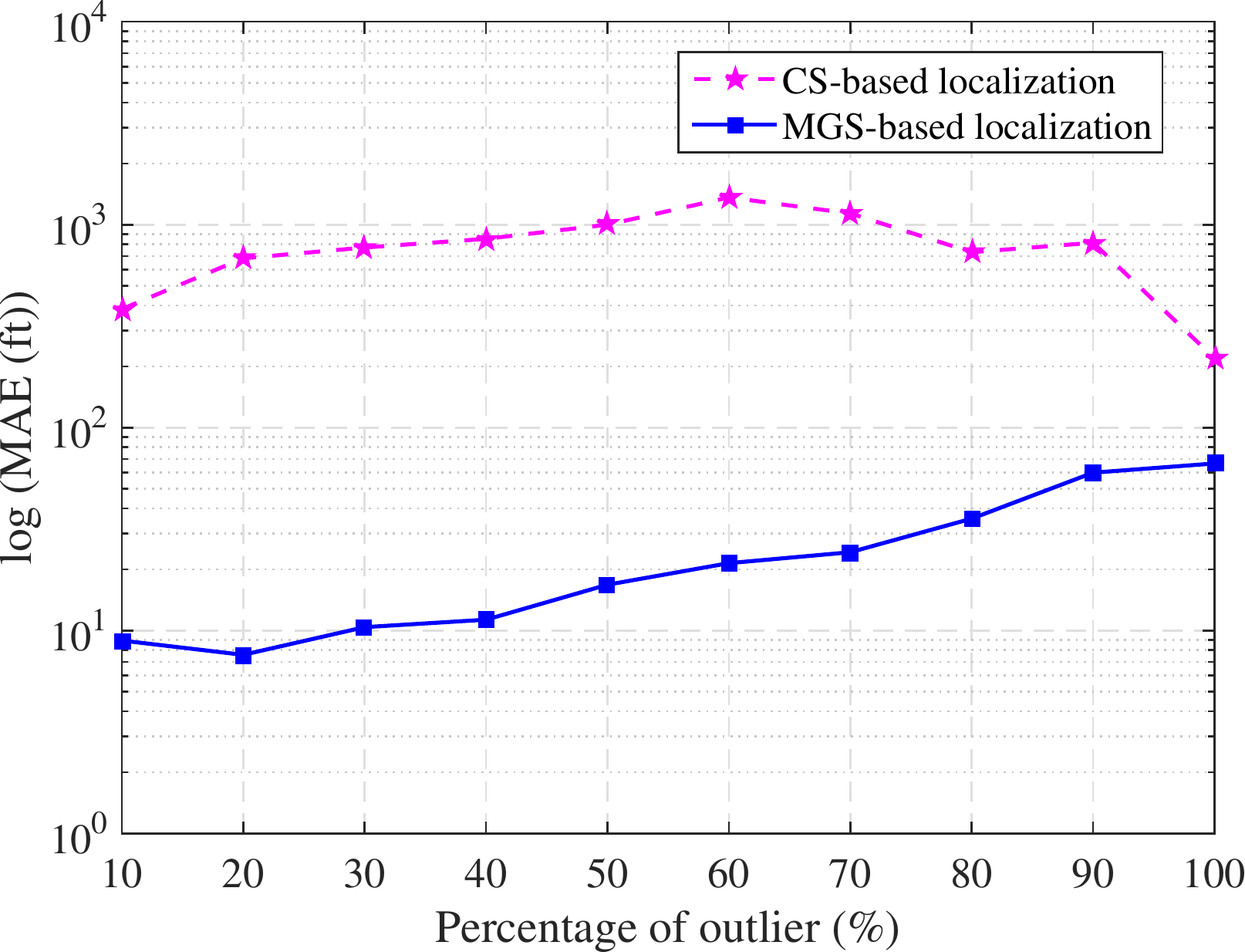}
                            \centering
                            \caption{The MAE for MGS-based and CS-based localization schemes with an increasing rate of APs that contain outlier. The MGS-based localization is robust to the outliers while the CS-based localization show very large localization errors even with a low percentage of outliers.}
                            \label{figure5}
                         \end{figure}
                         
                          \begin{figure}[t!]
                                                \includegraphics[scale=0.535]{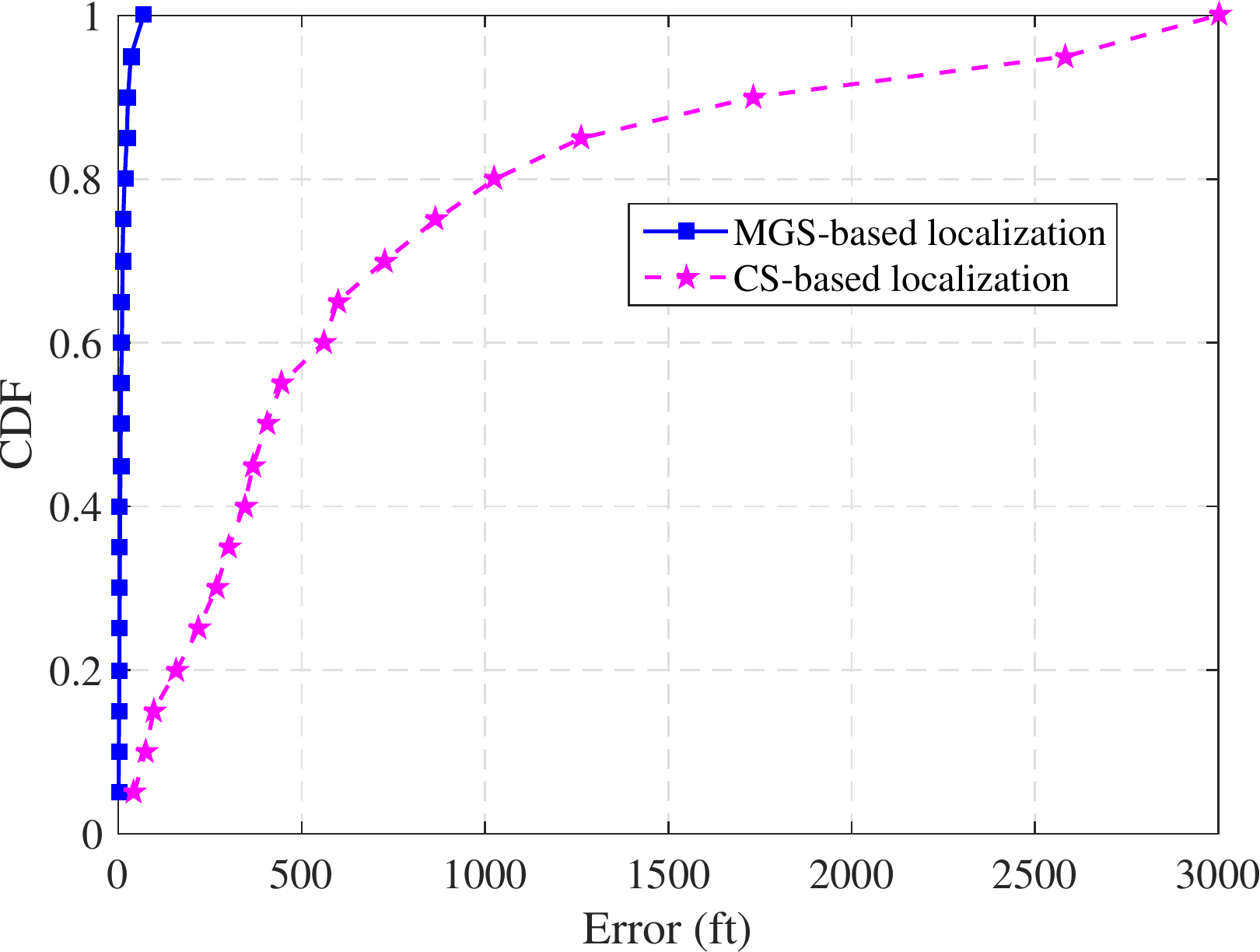}
                                                     \centering
                                                     \caption{The cumulative distribution localization error for MGS-based and CS-based localization schemes. The CS-based localization error are inordinate showing the sensitivity of the approach to the system malfunctioning. The MGS-based localization shows a great low distribution of errors.}
                                                     \label{figure8}
                                                  \end{figure}
\subsection {Experimental Results of MGS-based Localization Approach}
\par Fig. \ref{figure5} reports the MAE of the MGS-based and CS-based localization methods for an increasing rate of outliers in the APs. A total number of 21 APs have been used which may randomly be corrupted by outliers and the area is divided into $K=5$ groups. The figure clearly shows that the CS-based localization is sensitive to outliers even if a small number of APs are corrupted. Normally, in a  well-functioning system,  the percentage of the APs that experience outliers is low. However, the MGS-based localization shows a small increase in localization error when less than 9 APs are malfunctioning, indicating its robustness to outliers. The MGS-based approach is suitable when unprecedented events, like fire, flooding, and earthquake, occur  and a large portion of the infrastructure is damaged and emergency localization becomes crucial.
\par In order to evaluate how the errors are spread when APs contain outliers, the cumulative distribution function of errors for the MGS-based and CS-based localization approaches are shown in Fig. \ref{figure8} when 40 \% of the APs in the system do not function properly due to outliers. The results show the considerable difference between the two schemes, where the CS-based localization is completely misled. However, it is observed that the MGS-based localization has a maximum error of 67 ft.
 \begin{figure}[t!]
        \includegraphics[scale=0.47]{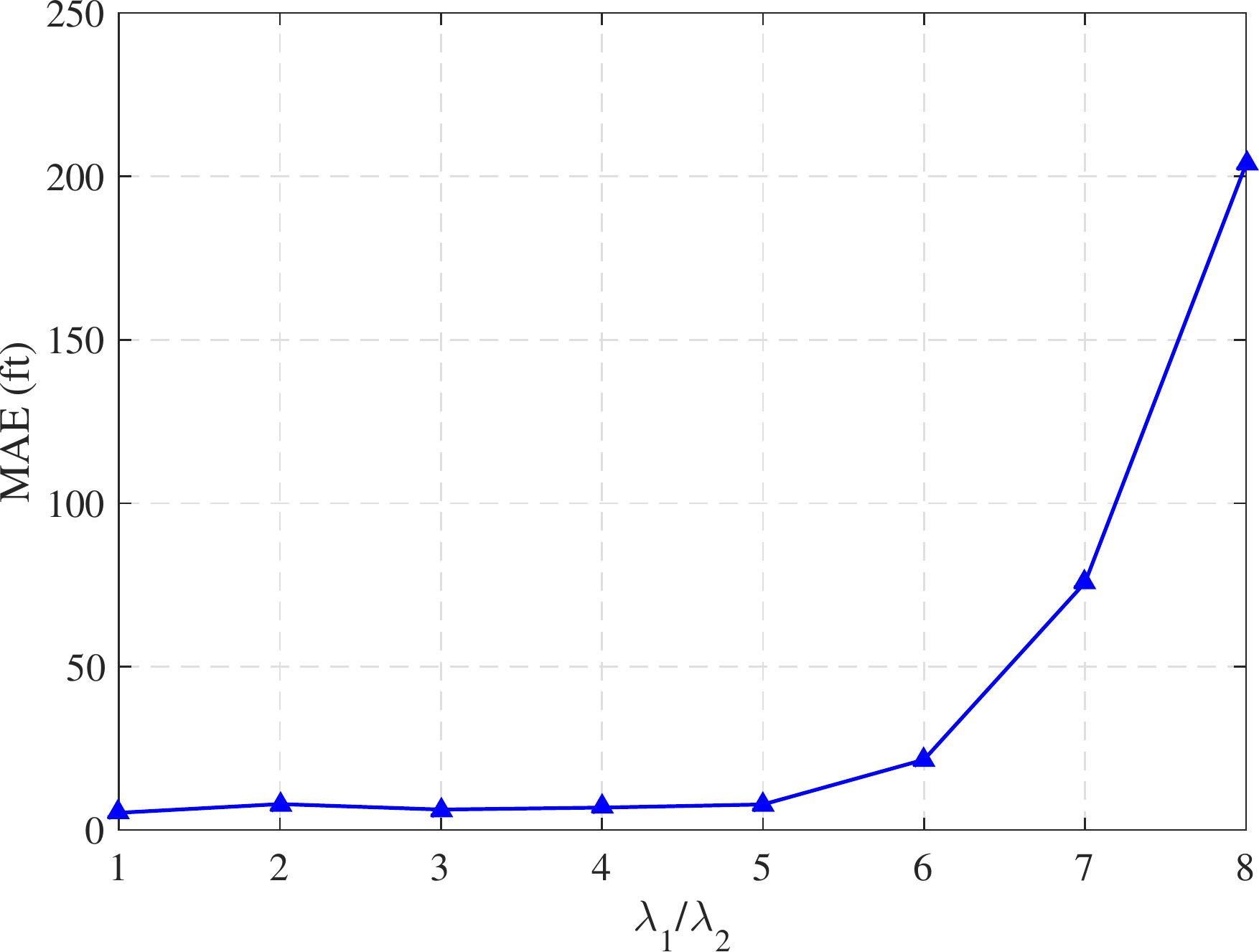}
        \centering
        \caption{The localization error of the GS-based approach for various ratios of $\lambda_{1}/\lambda_{2}$.}
         \label{figure11}    
 \end{figure}
\begin{figure}[t!]
        \includegraphics[scale=0.51]{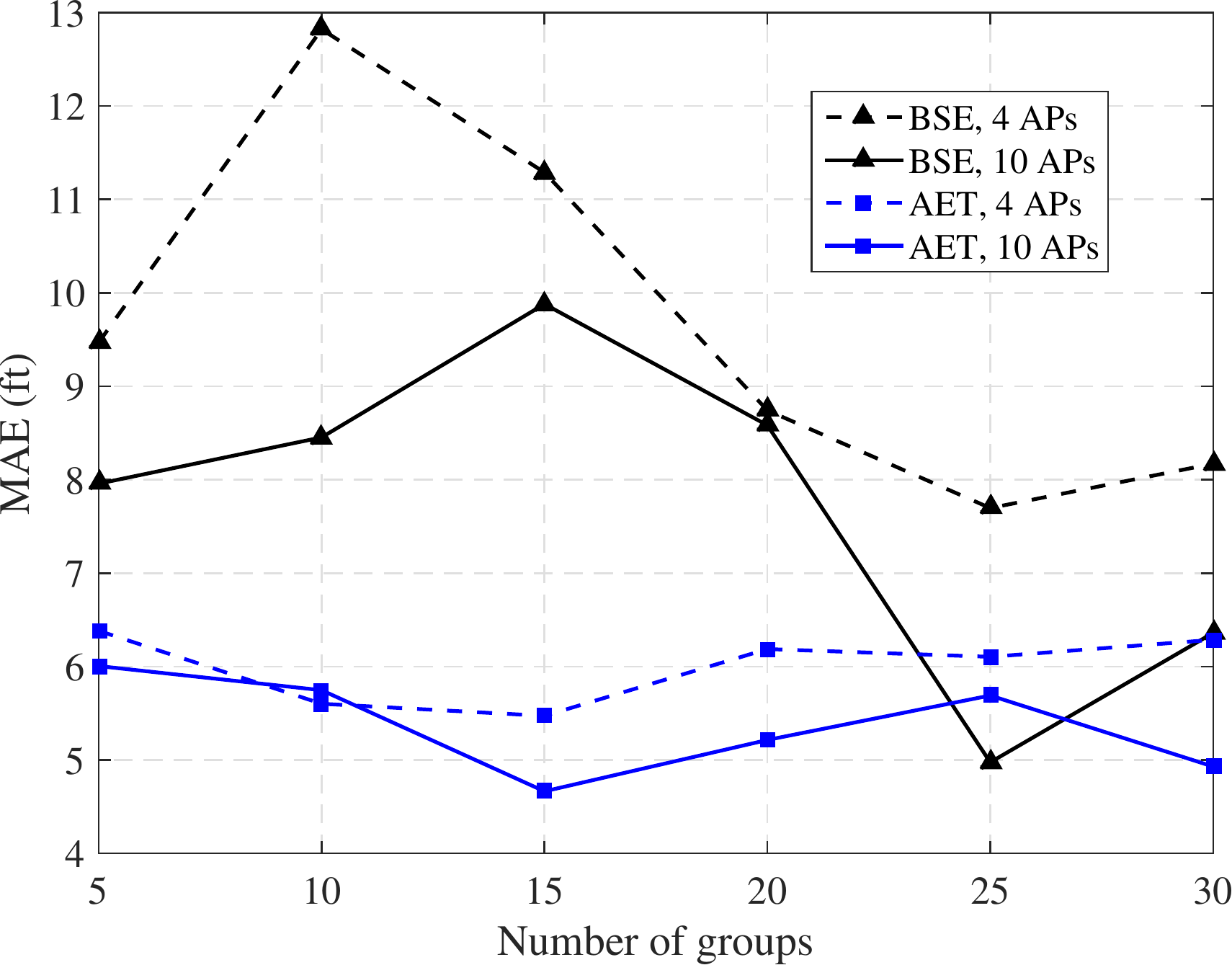}
        \centering
        \caption{The average localization error versus different number of groups for two different environments when 4 and 10 APs are used.}
         \label{figure12}  
 \end{figure}
 \subsection{Tuning Parameters}
\par There are different ways to tune the parameters $\lambda_{1} , \lambda_{2}$ for the GS-based and $\lambda_{1} , \lambda_{2},  \lambda_{3}$ for MGS-based localization schemes. One approach is the well-known Cross Validation (CV) \cite{r87}. The CV finds the Mean Square Error (MSE) of the residuals for a range of $\lambda_{i}, i=1,2,3$, and chooses the one with the least MSE. Another approach is to use training data and find the best values that minimize the localization error. The latter method, which has been used in this paper, is an easy and effective way for selecting these parameters. Only 10 training samples have been used for tuning. Fig. \ref{figure11} shows the localization error for an increasing $\lambda_{1}/\lambda_{2}$ when 12 APs have been used. The best parameter values must be calibrated for each building (e.g., at the fingerprinting stage). In the above results, we chose $\lambda_{1}/\lambda_{2}=0.5$ for GS-based localization and  $\lambda_{1}=\lambda_{2}=0.1 , \lambda_{3}=0.01$ for the MGS-based localization approach.
\par Since the GS-based localization needs to divide the area into groups and assign a weight to each, the number of groups also plays an important role in localization accuracy. To this end, an experiment is conducted to evaluate the  localization error for different number of groups and the results are shown in Fig. \ref{figure12} for AET and BSE building environments when 4 and 10 APs are used for localization.  The results illustrate that 15 groups are optimal to achieve the highest localization accuracy compared to 25 groups for BSE environment.  The BSE environment is structurally more complex than the AET environment and hence, it introduces a diverse set of RPs. However, it is clear that small localization errors for both environments are achieved which indicates the consistency of the method in different environments.
 \begin{figure}[t!]
        \includegraphics[scale=0.487]{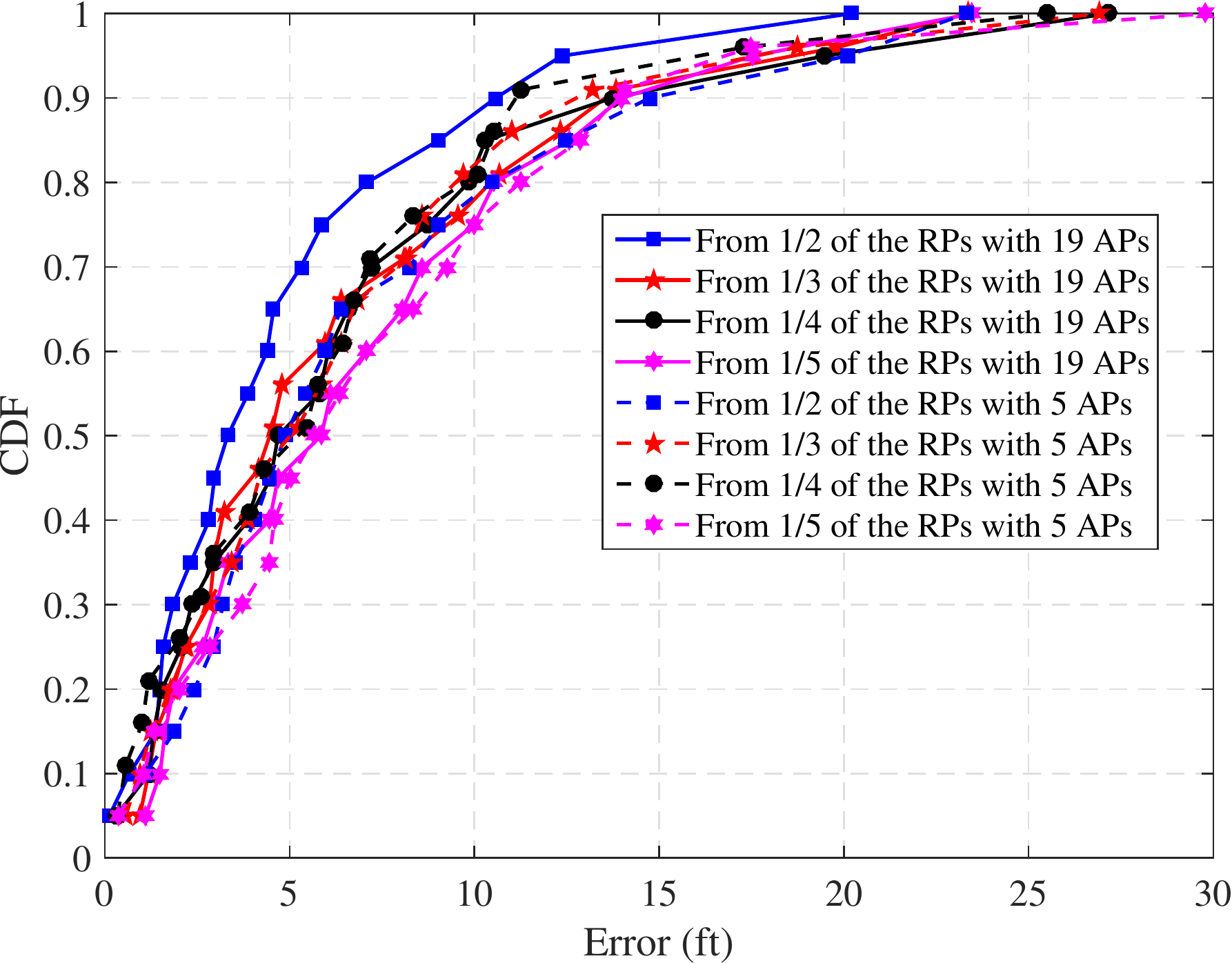}
        \centering
        \caption{The cumulative distribution of the localization error when $\frac{1}{2}$, $\frac{1}{3}$, $\frac{1}{4}$, $\frac{1}{5}$ of the RPs are used for radio map interpolation when 5 and 19 APs are used for localization.}
         \label{figure9}    
 \end{figure}
 \begin{center}
 \begin{threeparttable}[t]
 \caption{The Average Recovery Error of the Reconstructed Radio Map}
 \label{table1}
 \begin{tabular}{rlcc}
 \hline
 &   & Reconstruction Error (dBm) \\
 \hline
 &$1/2$  & $1.02$  \\
 &$1/3$ & $5.8$  \\
 &$1/4$ & $8.52$  \\
 &$1/5$ & $11.28$  \\
 \hline
 \end{tabular}
 \end{threeparttable}
 
 \end{center} 
 
\subsection {Experimental Results of LASSO-based Interpolation Scheme}
\par Furthermore, the LASSO-based approach of Section \ref{Radio map Interpolation using Sparse Recovery}, is utilized in the offline phase to reduce the number of fingerprints. For radio map interpolation, we chose 96, 64, 48, and 38 randomly selected RPs which lead to $\frac{1}{2}$, $\frac{1}{3}$, $\frac{1}{4}$, and $\frac{1}{5}$ of the total 192 RPs, respectively. The average recovery error of the radio map, which is the absolute difference between the actual and reconstructed radio map, called \textit{Reconstruction Error (RE)}, is computed as: 
\begin{equation}
\text{RE}=\frac{1}{N\times L} \sum_{i=1}^{L}\sum_{j=1}^{N} \left| \psi^{i}_{j}-\hat{\psi}^{i}_{j}\right|.
\end{equation}
The radio map reconstruction error for various ratios of selected RPs are shown in Table \ref{table1}. The reconstruction error when 1/2 of RPs is used for localization is 1.02 dB that is comparable to the error reported in \cite{r27}.

\par Fig. \ref{figure9} represents the results for the CDF of the localization error with the reconstructed radio map when 5 and 19 APs are selected. The proposed scheme is able to achieve a localization error of less than 10  ft for 88 \% of the time, when half of the fingerprints are interpolated from the actual RPs fingerprints and 19 APs are used. This only reduces to error of less than 10 ft for 80 \% . The results show that large fingerprinting cost and time savings can be achieved while tolerating a small increase in the localization error.
\par In sum, the previous results show two advantages of our approach. First, our scheme exhibits higher performance than the recent competing alternative. Second, the method can be utilized in emergency applications where the WLAN system is not functioning properly, and it is not clear which APs have problems or are unavailable.

\section{Conclusion}
\label{Conclusion}
In this paper, we have proposed three novel schemes for indoor WLAN fingerprinting localization, outlier detection, and radio map interpolation. The Group-Sparsity localization system does not rely on the typical coarse localization that other schemes require. Instead, the reference points are grouped in clusters, where each cluster participates in localization with its appropriate weight. Thus, the problem of wrong coarse localization has been avoided. The proposed approach uses convex optimization to localize the user. This minimization has three components that include the residuals between the radio map and the online measurements, the $\ell_{1}$-norm of the user's location for sparse recovery, and weighted $\ell_{2}$-norm of the groups of RPs to introduce the group sparsity. 
\par Since the AP online readings are prone to outliers due to unavailability or adversary attacks on APs, an outliers detection scheme also relying on sparse regression was developed. By explicitly modeling the outliers, the outliers are jointly estimated with the user's location vector.
\par To reduce the time-consuming and labor intensive radio map fingerprinting, a special form of Sparse Group Lasso regression was introduced which enables the fingerprints to be recorded at a coarse grid and radio map can be interpolated at finer grids.
\par The positioning system has been evaluated using real data from a high-commuting building that includes offices, labs, and a library. The experimental results demonstrate that the proposed scheme leads to substantial localization accuracy performance over the conventional fingerprinting methods.

\ifCLASSOPTIONcaptionsoff
  \newpage
\fi

\appendix
\section{Selecting RPs for Interpolation}
The selection of RPs for interpolation can also be rendered as sampling from all the available RPs and recording the respective fingerprints. This appendix uses the previous derivation to illustrate that regular sampling of the radio map lead to very inaccurate radio map interpolation, and advocates the use of random sampling.
\par In particular, the regular sampling of RPs can be thought as two consecutive procedures: 1) Compression (down-sampling) of the readings with a rate of $s$ for each single AP, $\boldsymbol \psi^{i}$, by $s$ RPs as
 \begin{equation}
 \begin{split}
\boldsymbol \psi^{i}_{down}[n]&=\psi^{i}_{sn} , \ \  n=1,\dots, \frac{N}{s} ,\\
  \forall i&=1,\dots,L
 \end{split}
 \end{equation}
 followed by an extension (up-sampling) with the rate of $s$,  which inserts $s$ zeros in between each RP of $\boldsymbol \psi^{i}_{down}[n]$
 \begin{equation}
 \label{eq26}
  \begin{split}
 \boldsymbol \psi^{i}_{down-up}[n]&= \begin{cases}
           \boldsymbol \psi^{i}_{down}[\frac{n}{s}] & \ \text{if} \ n=sk \ \ k=1,\dots, \frac{N}{s}\\
           0 & \text{otherwise}
       \end{cases} \ \ \ \\
     &= \begin{cases}
                  \boldsymbol \psi^{i}_{down} [n] & \ \text{if} \ n=sk \ \ k=1,\dots, \frac{N}{s}\\
                  0 & \text{otherwise}
              \end{cases} \ \ \ 
  \end{split}
  \end{equation}
 According to the above sampling procedure, each row of the sampling matrix $\mathbf{A}$ is a sparse vector that has only one nonzero value whose index shows the RP that we have recorded the fingerprints. 
 \par However, our results showed that the structure of periodic RP selection, in which $\mathbf{A}$ has periodic sampling structure, does not necessarily lead to a unique solution for $\boldsymbol \psi^{i}_{f}$ in \eqref{eq19} for all $s=1,2,\dots,N$. In particular, it is not hard to check that the Fourier transform of \eqref{eq26} [according to \eqref{eq17}] will satisfy the measurement equation \eqref{eq19} exactly. As such, it is likely that the outcome of the optimization \eqref{eq18} will be \eqref{eq26}, which is clearly a very inadequate interpolation of the original radio map. 
\par Therefore, the RPs whose fingerprints are engaged in interpolation should be selected randomly. In this case, the structure of the rows of $\mathbf{A}$ is randomly sparse.

\bibliographystyle{IEEEtran}
\bibliography{mybib}

\begin{thebibliography}{10}
\providecommand{\url}[1]{#1}
\csname url@samestyle\endcsname
\providecommand{\newblock}{\relax}
\providecommand{\bibinfo}[2]{#2}
\providecommand{\BIBentrySTDinterwordspacing}{\spaceskip=0pt\relax}
\providecommand{\BIBentryALTinterwordstretchfactor}{4}
\providecommand{\BIBentryALTinterwordspacing}{\spaceskip=\fontdimen2\font plus
\BIBentryALTinterwordstretchfactor\fontdimen3\font minus
  \fontdimen4\font\relax}
\providecommand{\BIBforeignlanguage}[2]{{%
\expandafter\ifx\csname l@#1\endcsname\relax
\typeout{** WARNING: IEEEtran.bst: No hyphenation pattern has been}%
\typeout{** loaded for the language `#1'. Using the pattern for}%
\typeout{** the default language instead.}%
\else
\language=\csname l@#1\endcsname
\fi
#2}}
\providecommand{\BIBdecl}{\relax}
\BIBdecl

\bibitem{1}
M.~McGuire, K.~Plataniotis, and A.~Venetsanopoulos, ``Data fusion of power and
  time measurements for mobile terminal location,'' \emph{IEEE Transactions on
  Mobile Computing}, vol.~4, no.~2, pp. 142--153, Mar. 2005.

\bibitem{2}
C.~Sytems, ``Wi-{F}i based real-time location tracking: Solutions and
  technology,'' CISCO Sytems, Tech. Rep., 2006.

\bibitem{3}
``Ekahau,'' http://www.ekahau.com, 2006.

\bibitem{4}
M.~Rodriguez, J.~Favela, E.~Martinez, and M.~Munoz, ``Location-aware access to
  hospital information and services,'' \emph{IEEE Transactions on Information
  Technology in Biomedicine}, vol.~8, no.~4, pp. 448--455, Dec. 2004.

\bibitem{5}
H.~Harroud, M.~Ahmed, and A.~Karmouch, ``Policy-driven personalized multimedia
  services for mobile users,'' \emph{IEEE Transactions on Mobile Computing},
  vol.~2, no.~1, pp. 16--24, Jan. 2003.

\bibitem{6}
R.~Muntz and C.~Pancake, ``Challenges in location-aware computing,'' \emph{IEEE
  Pervasive Computing}, vol.~2, no.~2, pp. 80--89, Apr. 2003.

\bibitem{7}
P.~Misra and P.~Enge, \emph{{Global Positioning System: Signals, Measurements,
  and Performance}}, 2nd~ed.\hskip 1em plus 0.5em minus 0.4em\relax
  Ganga-Jamuna Press, Lincoln MA, 2006.

\bibitem{58}
T.~D. Hodes, R.~H. Katz, E.~Servan-Schreiber, and L.~Rowe, ``Composable
  {A}d-hoc mobile services for universal interaction,'' in \emph{Proceedings of
  the 3rd Annual ACM/IEEE International Conference on Mobile Computing and
  Networking}, 1997, pp. 1--12.

\bibitem{r71}
D.~Pastina, F.~Colone, T.~Martelli, and P.~Falcone, ``Parasitic exploitation of
  {W}i-{F}i signals for indoor radar surveillance,'' \emph{IEEE Transactions on
  Vehicular Technology}, vol.~64, no.~4, pp. 1401--1415, Apr. 2015.

\bibitem{10}
U.~Bandara, M.~Hasegawa, M.~Inoue, H.~Morikawa, and T.~Aoyama, ``Design and
  implementation of a bluetooth signal strength based location sensing
  system,'' in \emph{IEEE Radio and Wireless Conference}, Sep. 2004, pp.
  319--322.

\bibitem{9}
R.~Want, A.~Hopper, V.~Falcao, and J.~Gibbons, ``The active badge location
  system,'' \emph{ACM Transactions on Information Systems}, vol.~10, no.~1, pp.
  91--102, 1992.

\bibitem{r81}
Y.-S. Kuo, P.~Pannuto, K.-J. Hsiao, and P.~Dutta, ``Luxapose: Indoor
  positioning with mobile phones and visible light,'' in \emph{Proceedings of
  the 20th Annual International Conference on Mobile Computing and Networking},
  2014, pp. 447--458.

\bibitem{r82}
Y.-C. Tung and K.~G. Shin, ``Echo{T}ag: Accurate infrastructure-free indoor
  location tagging with smartphones,'' in \emph{Proceedings of the 21st Annual
  International Conference on Mobile Computing and Networking}, 2015, pp.
  525--536.

\bibitem{r212}
S.~He and S.~H.~G. Chan, ``Wi-{F}i fingerprint-based indoor positioning: Recent
  advances and comparisons,'' \emph{IEEE Communications Surveys and Tutorials},
  vol.~18, no.~1, pp. 466--490, First Quarter 2016.

\bibitem{r215}
R.~Harle, ``A survey of indoor inertial positioning systems for pedestrians,''
  \emph{IEEE Communications Surveys and Tutorials}, vol.~15, no.~3, pp.
  1281--1293, Third Quarter 2013.

\bibitem{r62}
H.~Liu, H.~Darabi, P.~Banerjee, and J.~Liu, ``Survey of wireless indoor
  positioning techniques and systems,'' \emph{IEEE Transactions on Systems,
  Man, and Cybernetics, Part C: Applications and Reviews}, vol.~37, no.~6, pp.
  1067--1080, Nov 2007.

\bibitem{18}
M.~Youssef, A.~Agrawala, and A.~Udaya~Shankar, ``{WLAN} location determination
  via clustering and probability distributions,'' in \emph{Proceedings of the
  1st IEEE International Conference on Pervasive Computing and Communications},
  March 2003, pp. 143--150.

\bibitem{19}
A.~Kushki, K.~Plataniotis, and A.~Venetsanopoulos, ``Kernel-based positioning
  in wireless local area networks,'' \emph{IEEE Transactions on Mobile
  Computing}, vol.~6, no.~6, pp. 689--705, June 2007.

\bibitem{20}
Y.~Chen, Q.~Yang, J.~Yin, and X.~Chai, ``Power-efficient access-point selection
  for indoor location estimation,'' \emph{IEEE Transactions on Knowledge and
  Data Engineering}, vol.~18, no.~7, pp. 877--888, July 2006.

\bibitem{21}
A.~Kushki, K.~Plataniotis, A.~Venetsanopoulos, and C.~Regazzoni, ``Radio map
  fusion for indoor positioning in wireless local area networks,'' in \emph{in
  Proceedings of the 8th International Conference on Information Fusion},
  vol.~2, July 2005, pp. 8 pp.--.

\bibitem{r39}
A.~Tabibiazar and O.~Basir, ``Compressive sensing indoor localization,'' in
  \emph{2011 IEEE International Conference on Systems, Man, and Cybernetics
  (SMC)}, Oct 2011, pp. 1986--1991.

\bibitem{r27}
C.~Feng, W.~Au, S.~Valaee, and Z.~Tan, ``Received-signal-strength-based indoor
  positioning using compressive sensing,'' \emph{IEEE Transactions on Mobile
  Computing}, vol.~11, no.~12, pp. 1983--1993, Dec. 2012.

\bibitem{r33}
J.~Deng, Q.~Cui, X.~Zhang, and X.~Xu, ``Compressive sensing based indoor
  positioning with denosing and filtering in {LF} space,'' in \emph{23rd
  International Symposium on Personal Indoor and Mobile Radio Communications
  (PIMRC)}, Sep. 2012, pp. 2477--2482.

\bibitem{54}
S.-H. Fang, T.-N. Lin, and K.-C. Lee, ``A novel algorithm for multipath
  fingerprinting in indoor {WLAN} environments,'' \emph{IEEE Transactions on
  Wireless Communications}, vol.~7, no.~9, pp. 3579--3588, Sep. 2008.

\bibitem{55}
E.~Kupershtein, M.~Wax, and I.~Cohen, ``Single-site emitter localization via
  multipath fingerprinting,'' \emph{IEEE Transactions on Signal Processing},
  vol.~61, no.~1, pp. 10--21, Jan. 2013.

\bibitem{r70}
M.~Bshara, U.~Orguner, F.~Gustafsson, and L.~Van~Biesen, ``Fingerprinting
  localization in wireless networks based on received-signal-strength
  measurements: A case study on {WiMAX} networks,'' \emph{IEEE Transactions on
  Vehicular Technology}, vol.~59, no.~1, pp. 283--294, Jan. 2010.

\bibitem{r72}
T.-N. Lin, S.-H. Fang, W.-H. Tseng, C.-W. Lee, and J.-W. Hsieh, ``A
  group-discrimination-based access point selection for {WLAN} fingerprinting
  localization,'' \emph{IEEE Transactions on Vehicular Technology}, vol.~63,
  no.~8, pp. 3967--3976, Oct. 2014.

\bibitem{r73}
J.~Hong and T.~Ohtsuki, ``Signal eigenvector-based device-free passive
  localization using array sensor,'' \emph{IEEE Transactions on Vehicular
  Technology}, vol.~64, no.~4, pp. 1354--1363, Apr. 2015.

\bibitem{r74}
S.-H. Fang and C.-H. Wang, ``A dynamic hybrid projection approach for improved
  {W}i-{F}i location fingerprinting,'' \emph{IEEE Transactions on Vehicular
  Technology}, vol.~60, no.~3, pp. 1037--1044, Mar. 2011.

\bibitem{56}
K.~Kaemarungsi and P.~Krishnamurthy, ``Modeling of indoor positioning systems
  based on location fingerprinting,'' in \emph{Proceedings of the 23rd Annual
  Joint Conference of the IEEE Computer and Communications Societies}, vol.~2,
  Mar. 2004, pp. 1012--1022 vol.2.

\bibitem{r195}
Z.~Xiao, H.~Wen, A.~Markham, and N.~Trigoni, ``Lightweight map matching for
  indoor localisation using conditional random fields,'' in \emph{Proceedings
  of the 13th International Symposium on Information Processing in Sensor
  Networks}, Apr. 2014, pp. 131--142.

\bibitem{r230}
P.~Mirowski, H.~Steck, P.~Whiting, R.~Palaniappan, M.~MacDonald, and T.~K. Ho,
  ``{KL}-divergence kernel regression for non-gaussian fingerprint based
  localization,'' in \emph{Proceedings of the International Conference on
  Indoor Positioning and Indoor Navigation (IPIN)}, Sep. 2011, pp. 1--10.

\bibitem{r234}
S.~He and S.~H.~G. Chan, ``Tilejunction: Mitigating signal noise for
  fingerprint-based indoor localization,'' \emph{IEEE Transactions on Mobile
  Computing}, vol.~15, no.~6, pp. 1554--1568, June 2016.

\bibitem{r235}
------, ``Sectjunction: Wi-fi indoor localization based on junction of signal
  sectors,'' in \emph{Proceedings of the IEEE International Conference on
  Communications (ICC)}, June 2014, pp. 2605--2610.

\bibitem{r237}
S.~He, T.~Hu, and S.-H.~G. Chan, ``Contour-based trilateration for indoor
  fingerprinting localization,'' in \emph{Proceedings of the 13th ACM
  Conference on Embedded Networked Sensor Systems}, 2015, pp. 225--238.

\bibitem{r68}
M.~Wang and C.~Zhang, ``Residual ranking: A robust access-point selection
  strategy for indoor location tracking,'' in \emph{Proceedings of the IEEE
  International Conference on Systems, Man and Cybernetics}, Oct. 2009, pp.
  5035--5040.

\bibitem{r75}
V.~S. Feng, T.~C. Wang, S.~Y. Chang, and H.-P. Ma, ``Location estimation in
  indoor wireless networks by hierarchical support vector machines with fast
  learning algorithm,'' in \emph{International Conference on System Science and
  Engineering (ICSSE)}, July 2010, pp. 321--326.

\bibitem{r76}
W.~Yanhua, W.~Dongli, and Z.~Yan, ``Axial decoupled {LS-SVM}s for indoor
  positioning using {RSS} fingerprints,'' in \emph{Proceedings of the 34th
  Chinese Control Conference (CCC)}, July 2015, pp. 3920--3925.

\bibitem{8}
P.~Bahl and V.~N. Padmanabhan, ``{RADAR}: An in-building {RF}-based user
  location and tracking system,'' in \emph{Proceedings of the IEEE
  International Conference on Computer Communications}, 2000, pp. 775--784.

\bibitem{r64}
J.~Ma, X.~Li, X.~Tao, and J.~Lu, ``Cluster filtered {KNN}: A {WLAN}-based
  indoor positioning scheme,'' in \emph{International Symposium on a World of
  Wireless, Mobile and Multimedia Networks}, June 2008, pp. 1--8.

\bibitem{r77}
M.~Dakkak, B.~Daachi, A.~Nakib, and P.~Siarry, ``Multi-layer perceptron neural
  network and nearest neighbor approaches for indoor localization,'' in
  \emph{Proceedings of the IEEE International Conference on Systems, Man and
  Cybernetics (SMC)}, Oct. 2014, pp. 1366--1373.

\bibitem{r78}
B.~Li, J.~Salter, A.~G. Dempster, and C.~Rizos, ``Indoor positioning techniques
  based on wireless {LAN},'' in \emph{Proceedings of the IEEE Ineternational
  Conference on Wireless Broadband and Ultra Wideband Communications}, pp.
  13--16.

\bibitem{31}
E.~Candes and J.~Romberg, ``Sparsity and incoherence in compressive sampling,''
  2006.

\bibitem{32}
E.~Candes and M.~Wakin, ``An introduction to compressive sampling,''
  \emph{Signal Processing Magazine, IEEE}, vol.~25, no.~2, pp. 21--30, March
  2008.

\bibitem{r93}
J.~Deng, Q.~Cui, and X.~Zhang, ``Data pre-processing in compressive sensing
  based indoor fingerprinting positioning,'' \emph{International Journal of
  Wireless Information Networks}, vol.~20, no.~4, pp. 256--267, 2013.

\bibitem{r83}
A.~T. Mariakakis, S.~Sen, J.~Lee, and K.-H. Kim, ``{SAIL}: Single access
  point-based indoor localization,'' in \emph{Proceedings of the 12th Annual
  International Conference on Mobile Systems, Applications, and Services},
  2014, pp. 315--328.

\bibitem{r177}
W.~Sun, J.~Liu, C.~Wu, Z.~Yang, X.~Zhang, and Y.~Liu, ``Mo{L}oc: On
  distinguishing fingerprint twins,'' in \emph{Proceedings of the 33rd IEEE
  International Conference on Distributed Computing Systems (ICDCS)}, July
  2013, pp. 226--235.

\bibitem{r197}
H.~Wen, Z.~Xiao, N.~Trigoni, and P.~Blunsom, ``On assessing the accuracy of
  positioning systems in indoor environments,'' in \emph{Proceedings of 10th
  European Conference on Wireless Sensor Networks}.\hskip 1em plus 0.5em minus
  0.4em\relax Springer Berlin Heidelberg, 2013, pp. 1--17.

\bibitem{r199}
A.~W.~S. Au, C.~Feng, S.~Valaee, S.~Reyes, S.~Sorour, S.~N. Markowitz, D.~Gold,
  K.~Gordon, and M.~Eizenman, ``Indoor tracking and navigation using received
  signal strength and compressive sensing on a mobile device,'' \emph{IEEE
  Transactions on Mobile Computing}, vol.~12, no.~10, pp. 2050--2062, Oct.
  2013.

\bibitem{r200}
A.~Kushki, K.~Plataniotis, and A.~Venetsanopoulos, ``Location tracking in
  wireless local area networks with adaptive radio maps,'' in \emph{Proceedings
  of the IEEE International Conference on Acoustics, Speech and Signal
  Processing}, vol.~5, May 2006.

\bibitem{r90}
M.~M. Atia, A.~Noureldin, and M.~J. Korenberg, ``Dynamic online-calibrated
  radio maps for indoor positioning in wireless local area networks,''
  \emph{IEEE Transactions on Mobile Computing}, vol.~12, no.~9, pp. 1774--1787,
  Sep. 2013.

\bibitem{r91}
C.~Koweerawong, K.~Wipusitwarakun, and K.~Kaemarungsi, ``Indoor localization
  improvement via adaptive {RSS} fingerprinting database,'' in
  \emph{Proceedings of The International Conference on Information Networking},
  Jan. 2013, pp. 412--416.

\bibitem{r92}
R.~Gao, M.~Zhao, T.~Ye, F.~Ye, Y.~Wang, K.~Bian, T.~Wang, and X.~Li, ``Jigsaw:
  Indoor floor plan reconstruction via mobile crowdsensing,'' in
  \emph{Proceedings of the 20th Annual International Conference on Mobile
  Computing and Networking}, 2014, pp. 249--260.

\bibitem{53}
Y.~Chen, W.~Trappe, and R.~Martin, ``{ADLS}: Attack detection for wireless
  localization using least squares,'' in \emph{Proceedings of the 5th Annual
  IEEE International Conference on Pervasive Computing and Communications
  Workshops}, Mar. 2007, pp. 610--613.

\bibitem{r80}
K.~Kaemarungsi and P.~Krishnamurthy, ``Properties of indoor received signal
  strength for {WLAN} location fingerprinting.''\hskip 1em plus 0.5em minus
  0.4em\relax IEEE Computer Society, 2004, pp. 14--23.

\bibitem{46}
R.~O. Duda, P.~E. Hart, and D.~G. Stork, \emph{Pattern Classification (2nd
  Edition)}.\hskip 1em plus 0.5em minus 0.4em\relax Wiley-Interscience, 2000.

\bibitem{51}
V.~Kekatos and G.~Giannakis, ``From sparse signals to sparse residuals for
  robust sensing,'' \emph{IEEE Transactions on Signal Processing}, vol.~59,
  no.~7, pp. 3355--3368, July 2011.

\bibitem{60}
N.~Simon, J.~Friedman, T.~Hastie, and R.~Tibshirani, ``A sparse-group lasso,''
  \emph{Journal of Computational and Graphical Statistics}, 2013.

\bibitem{61}
J.~Liu, S.~Ji, and J.~Ye, \emph{{SLEP}: Sparse Learning with Efficient
  Projections}, Arizona State University, 2009.

\bibitem{cvx}
I.~CVX~Research, ``{CVX}: Matlab software for disciplined convex programming,
  version 2.0,'' \url{http://cvxr.com/cvx}, Aug. 2012.

\bibitem{gb08}
M.~Grant and S.~Boyd, ``Graph implementations for nonsmooth convex programs,''
  in \emph{Recent Advances in Learning and Control}, ser. Lecture Notes in
  Control and Information Sciences, V.~Blondel, S.~Boyd, and H.~Kimura,
  Eds.\hskip 1em plus 0.5em minus 0.4em\relax Springer-Verlag Limited, 2008,
  pp. 95--110, \url{http://stanford.edu/~boyd/graph_dcp.html}.

\bibitem{r86}
Z.~Yang, C.~Wu, and Y.~Liu, ``Locating in fingerprint space: Wireless indoor
  localization with little human intervention,'' in \emph{Proceedings of the
  18th Annual International Conference on Mobile Computing and Networking},
  2012, pp. 269--280.

\bibitem{r85}
R.~Nandakumar, K.~K. Chintalapudi, and V.~N. Padmanabhan, ``Centaur: Locating
  devices in an office environment,'' in \emph{Proceedings of the 18th Annual
  International Conference on Mobile Computing and Networking}, 2012, pp.
  281--292.

\bibitem{r87}
T.~Hastie, R.~Tibshirani, and J.~Friedman, \emph{The Elements of Statistical
  Learning}, 2001.

\end{thebibliography}

\newpage
\begin{IEEEbiography}[
{\includegraphics[width=1in,height=1.25in,clip,keepaspectratio, angle =0 ]{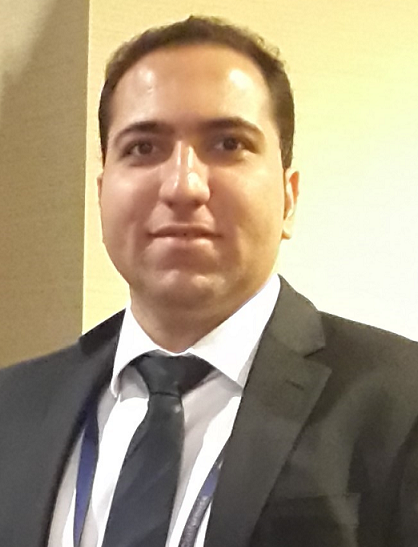}}
]{Ali Khalajmehrabadi} received his B.Sc. from Babol Noshirvani University of Technology, Iran, in 2010 and M.Sc. from University Technology Malaysia (UTM), Malaysia, in 2012 with the best graduate student award. He is currently a Ph.D. candidate (with specialization in Communications) in the Department of Electrical and Computer Engineering, the University of Texas at San Antonio (UTSA). His research interests include collaborative localization, indoor localization and navigation systems, and Global Positioning System (GPS). He is a student member of IEEE. 
\end{IEEEbiography}
\vspace{-20mm}
\begin{IEEEbiography}[
{\includegraphics[width=1in,height=1.25in,clip,keepaspectratio, angle =0 ]{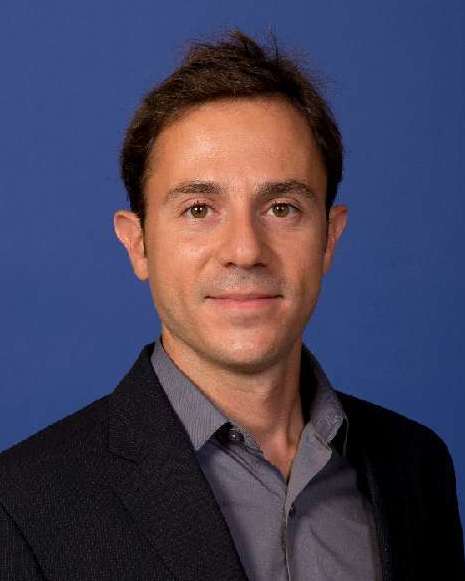}}
]{Nikolaos Gatsis}
 received the Diploma degree in Electrical and Computer Engineering from the University of Patras, Greece, in 2005 with honors. He received the M.Sc. degree in Electrical Engineering in 2010, and the Ph.D. degree in Electrical Engineering with minor in Mathematics in 2012, both from the University of Minnesota. He is currently an Assistant Professor with the Department of Electrical and Computer Engineering at the University of Texas at San Antonio. His research interests lie in the areas of smart power grids, communication networks, and cyberphysical systems, with an emphasis on optimal resource management. Prof. Gatsis co-organized symposia in the area of Smart Grids in IEEE GlobalSIP 2015 and IEEE GlobalSIP 2016. He also served as a Technical Program Committee member for symposia in IEEE SmartGridComm from to 2013 through 2016, and in GLOBECOM 2015.
\end{IEEEbiography}
\vspace{-20mm}

\begin{IEEEbiography} [

{\includegraphics[width=1in,height=1.25in,clip,keepaspectratio, angle =0 ]{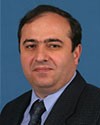}}
]{David Akopian} (M’02-SM’04) is a Professor at the University of Texas at San Antonio (UTSA). Prior to joining UTSA he was a Senior Research Engineer and Specialist with Nokia Corporation from 1999 to 2003. From 1993 to 1999 he was a researcher and instructor at the Tampere University of Technology, Finland, where he received his Ph.D. degree in electrical engineering in 1997. Dr. Akopian’s current research interests include digital signal processing algorithms for communication and navigation receivers, positioning, dedicated hardware architectures and platforms for software defined radio and communication technologies for healthcare applications. He authored and co-authored more than 30 patents and 140 publications. He served in organizing and program committees of many IEEE conferences and co-chairs annual SPIE Multimedia on Mobile Devices conference. His research has been supported by National Science Foundation, National Institutes of Health, USAF, US Navy and Texas foundations.
\end{IEEEbiography}

\end{document}